\documentclass{article}
\usepackage{authblk}
\usepackage{setspace}
\usepackage[margin=1.25in]{geometry}

\usepackage{mathptmx}       
\usepackage{helvet}         
\usepackage{courier}        
\usepackage{amsmath}
\usepackage{amsfonts}
\usepackage{caption}
\usepackage{subcaption}
\usepackage{makeidx}         
\usepackage{graphicx}        
\usepackage{multicol}        
\usepackage[bottom]{footmisc}
\usepackage{bm}

\usepackage[english]{babel}
\usepackage[utf8x]{inputenc}
\usepackage{algorithm,algorithmic}

\newcommand{\bs}[1]{\boldsymbol{#1}}
\newcommand{\eqn}[1]{Eqn.~(\ref{#1})}
\newcommand{\vek}[1]{\mathbf{#1}}
\newcommand{\tvek}[1]{\tilde{\mathbf{#1}}}

\newcommand{\hvek}[1]{\hat{\mathbf{#1}}}

\newcommand{\fig}[1]{Figure~\ref{#1}}
\newcommand{\alg}[1]{Algorithm~\ref{#1}}

\makeindex             

\begin{document}

\title{A robust algorithm for computational floating body dynamics}

\author[1,3*]{J. Roenby}
\author[2]{S. Aliyar}
\author[2]{H. Bredmose}

\affil[1]{Stromning Aps, Luftmarinegade 62, 1432 K\o benhavn K, DK}
\affil[2]{Department of Wind and Energy Systems, Technical University of Denmark, Nils Koppels Alle, Kgs. Lyngby 2800, DK}
\affil[3]{Department of Science and Environment, Roskilde University, Universitetsvej 1, 4000 Roskilde, DK}
\affil[*]{johan@ruc.dk}


\maketitle

\abstract{
We present a non--iterative algorithm, FloatStepper, for coupling the motion of a rigid body and an incompressible fluid in computational fluid dynamics (CFD) simulations. The purpose of the algorithm is to remove the so-called added mass instability problem, which may arise when a light floating body interacts with a heavy fluid. The idea underlying the presented coupling method is to precede every computational time step by a series of prescribed probe body motions in which the fluid response is determined, thus revealing the decomposition of the net force and torque into two components: 1) An added mass contribution proportional to the instantaneous body acceleration, and 2) all other forces and torques. The algorithm is implemented and released as an open source extension module to the widely used CFD toolbox, OpenFOAM, as an alternative to the existing body motion solvers.
The accuracy of the algorithm is investigated with several single-phase and two-phase flow benchmark cases. The benchmarks demonstrate excellent stability properties, allowing simulations even with massless bodies. They also highlight aspects of the implementation, such as the mesh motion method, where more work is needed to further enhance the flexibility and predictive capabilities of the code.}

\section{Introduction}
Accurate modelling of floating body motion is important for the design of offshore structures. This requires a robust numerical approach to both free surface flow and wave-structure interaction. While many basic response effects are well described by linear and second-order radiation-diffraction theory \cite{newmanMarineHydrodynamics}, this relies on the assumption of small wave steepness and small body motion. To describe response effects from large waves, slamming from breaking waves, green water flow on the structure and viscous damping, more accurate modelling is needed.

During the last decades, Computational Fluid Dynamics (CFD) for free surface flow has been matured to a level where it is now a viable solution for engineering calculation of design wave events. Among the various CFD methodologies, the Finite Volume Method (FVM) in combination with the Volume of Fluid technique (VoF) for the free surface treatment \cite{Hirt_1981_VOF}, has shown robust performance with ability to calculate breaking wave loads on e.g. monopiles, see for example \cite{Christensen2005_Extreme, Bredmose2006_NumericalReproduction, BredmoseAndJacobsen2010_Breaking}. Several publications have shown good comparisons to measured force and pressure for such breaking wave impacts, e.g. \cite{Ghadirian2019_Secondary}. Compared to potential flow solvers, CFD allows the fluid topology to change, for example through release of droplets. Much research effort has gone into the detailed VoF schemes to avoid numerical smearing of the air-water interface. Well--known methods after the original paper of Hirt \& Nicols \cite{Hirt_1981_VOF} include the works in \cite{Ubbink1999,deshpandeEvaluatingPerformanceTwophase2012}, as well as the isoAdvector scheme \cite{roenbyComputationalMethodSharp2016}, which can be applied on unstructured meshes.

Given the successful results for fixed structures, application of finite volume CFD to floating body motion appears to be a natural next development.  In principle, the body motion can be treated by calculation of its acceleration in each time step through Newtons second law, with input of the integrated fluid pressure as a force on the right hand side. Several studies with FVM-VoF based floating body CFD have been published in recent years, especially  within ship motion, wave energy generation and floating wind turbine motion see e.g.  \cite{Schmitt2015, Sarlak2018_CFD, Wang_SJTU_2019, Begovic_SurfRiding_2020} and \cite{Wangetal2022}. A thorough review on various solver types is given by Windt et al. \cite{Windt2018_Review}. Further, Ransley et al. \cite{Ransley2020_Blind} presented a comparative study for the response of focused wave groups for a hemispherical-bottomed buoy and a truncated cylinder with a cylindrical moon-pool with both potential flow solvers and CFD. A straight-forward implementation of the above steps, however, have shown to be unstable for bodies with low structural mass. 
The inherent problem is that acceleration of the body requires simultaneous acceleration of the surrounding fluid, such that extra fluid mass must be added to the body mass to achieve the truly needed force through Newtons second law. This fluid mass is denoted 'added mass' and is generally a $6\times 6$ matrix that depends on the instantaneous fluid topology. 

The need for a proper treatment of added mass in floating body CFD has been discussed already by S\"oding \cite{soding2001integrate} in a conference paper, that appears to have only little recognition. Bettle \cite{bettleUnsteadyComputationalFluid2012} discussed the stability problem in the context of CFD for submarine maneuvering and devised a coupling algorithm with iterations between body and fluid motion. A floating body solver along the lines of Bettle's work is found in the widely used open source CFD code, OpenFOAM. This was improved in the work of Dunbar et al. \cite{dunbarDevelopmentValidationTightly2015} and Chow et al. \cite{chowStronglyCoupledPartitioned2016} using dynamic relaxation techniques, and by Bruinsma et al. \cite{Brunisma2018_Validation} who stabilised solutions by relaxing fluid pressure in the iteration loop. The latter concluded that more work is needed to achieve a robust solution for the added mass problem, since the stabilization techniques lead to larger computational effort. Further steps in the solution of the added mass problem have been proposed by Devolder et al. \cite{devolderAcceleratedNumericalSimulations2019} in terms of an acceleration technique for the added mass iterations with 1 degree of freedom (DoF), and by Veldman et al. \cite{Veldman2016_FreeSurfaceFloating}, in terms of an approximate initial added mass term.

While the iterative methods can give accurate results upon convergence, a robust solution of the equations of motion, requires separation of the added mass force from the overall fluid force lumping the body and added mass together to properly isolate the acceleration in the force equation. This is the core idea of the present paper. We develop an algorithm, where the added mass is determined explicitly in each time step and thus allows an accurate and direct calculation of the true body acceleration without the need for outer iterations. We implement the algorithm and demonstrate its robustness through a series of numerical experiments.

It is worth to note, that the idea of an explicit added mass matrix in floating body modelling has been presented also by other researchers. In this respect, our approach has strong similarities to the algorithm outlined by S\"oding \cite{soding2001integrate} and applied by Shigonov et al \cite{Shigonov_2001} for aircraft landing on water in three degrees of freedom. In Söding's paper, the need for an explicit added mass matrix for numerical stability is explained and an iteration based method for its determination is formulated. In a study by Meyer et al. \cite{Meyer2017_NewAdjustmentFree} this algorithm was applied to calculate the motion of a yacht in head waves. To the best of our knowledge, no thorough demonstration of the stability properties exists in the literature, and no open source CFD implementation is available to the scientific community.

The contribution of the present work is to develop and implement in a fully parallelised unstructured FVM CFD code, OpenFOAM, a 6--DoF added mass aware algorithm which in contrast to earlier works is free of outer iteration loops. We analyse and demonstrate its stability properties and validate it against five test cases. Two of these have analytical solutions, a rising disc and a wiggling ellipse, and two contain comparison to experimental results, namely a freely floating box and a moored box in regular waves. Our hope is that the new robust method, and the release of the implementation as an OpenFOAM extension module  \cite{roenby_2023_8146516}, will provide a simpler approach to CFD simulations of floating body problems.


\section{The added mass instability problem}
We consider the numerical coupling of a floating rigid body with a surrounding incompressible fluid. 
The fluid may either be a single fluid or two immiscible fluids separated by a sharp fluid interface. 
It may either be inviscid or viscous with slip or no-slip boundary condition on the body so the instantaneous distribution of fluid pressure –- and possibly shear stress –- exerts a force and a torque on the body.

When an external force, $\vek F$, such as gravity or a spring works on a rigid body immersed in fluid, the body will accelerate and so must the surrounding fluid to accommodate the body displacement.
In a small time step, $\delta t$, the momentum of the body-fluid system will increase by $\vek F\delta t$ and from a numerical point of view the difficulty is that we do not know a priori how this increase in system momentum is distributed between the fluid and the body. 
It is the task of a coupling algorithm to find this distribution and advance the system accordingly.
\begin{figure}[tb]
    \centering
	\includegraphics[width=.9\textwidth]{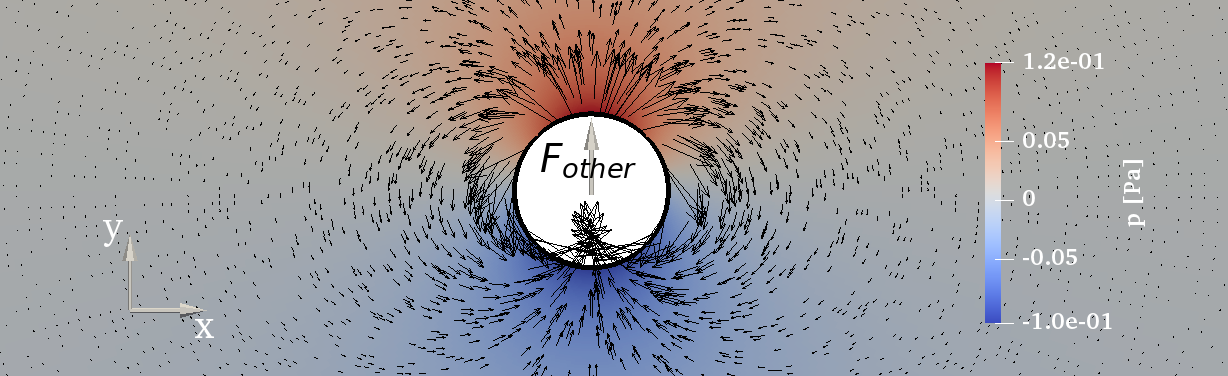}
    \caption{Pressure and velocity field around circular body exposed to a force along the $y$-axis.\label{fig:risingCircle}}
\end{figure}

For illustrative purposes let us consider a rigid cylinder floating in 2D ideal fluid of uniform mass density $\rho_f$, as shown in \fig{fig:risingCircle}. The body has radius $R$ and uniform mass density $\rho_b$, and hence mass (per unit length) $m_b = \rho_b\pi R^2$. 
If the body is exposed to a net force, $F$, along the $y$-axis, this can be written as
\begin{equation}\label{eq:Fdivision}
	F = F_\text{other} - m_a a,
\end{equation}
where $m_a$ is the added mass of the body, $a$ is the instantaneous $y$-acceleration of the body, and $F_\text{other}$ represents all other forces on the body (gravity, buoyancy, mooring lines etc.). In this simple example, the added mass is known a priori to be  $m_a = \rho_f \pi R^2$. Equating the total force, $F$, to $m_b a$, and isolating $a$, we get the body acceleration,
\begin{equation}\label{eq:accel}
	a = \frac{F_\textrm{other}}{m_a + m_b}.
\end{equation}
In computational fluid dynamics simulations we often do not know $m_a$ and/or $F_\text{other}$. 
Therefore, in partitioned coupling algorithms, we typically resort to iteration between
\begin{enumerate}
\item Calculating the body acceleration as $a = F/m_b$ (or some relaxed variant of this), where $F$ includes the force from the surrounding fluid flow, and 
\item Calculating the fluid flow and force on the body, $F$, resulting from the body acceleration, $a$.
\end{enumerate}
The hope is that iterations between these two steps will eventually converge to the physically correct body acceleration and fluid response, which is assumed to be reached when reiteration no longer changes the results (to within a tolerance).
It is, however, well--known that this iterative procedure is unstable when the body mass is smaller than the added mass \cite{causinAddedmassEffectDesign2005}.

Let us first consider a loose body--fluid coupling algorithm without any iterations. Each time step contains a single update of the body state followed by an update of the fluid state. We assume the only force on our circular body is gravity, $\vek g = -g\hvek y$, and so $F_\textrm{other}$ is constant in time.
The body state is then represented by the body position and velocity $(\vek x_b, \vek v_b)$, here restricted to motion along the $y$-axis. 
The fluid state is represented by the velocity field and pressure field $(\vek u, p)$. The algorithm could look as sketched in \alg{alg:looseCoupling}.
\begin{algorithm}[h!]
\caption{A simple loose coupling algorithm.}
\label{alg:looseCoupling}
\begin{algorithmic}[1]
\STATE Initialise body state, $(\vek x_b, \vek v_b)$, and fluid state, $(\vek u, p)$\label{stp:sysInit}.
\STATE Increment time by $\Delta t$\label{stp:+dt}.
\STATE Calculate the net force, $\vek F$, on the body including $p$ integrated over its surface
\STATE Calculate the body acceleration using Newton’s 2nd law: $\vek a = \vek F/m_b$\label{stp:calca}
\STATE Numerically integrate $\vek a$ to get the updated velocity $\vek v_b^\text{new}$, and position $\vek x_b^\text{new}$.
\STATE Update body state, $(\vek x_b, \vek v_b) = (\vek x_b^\text{new}, \vek v_b^\text{new})$, and update mesh accordingly. 
\STATE Update fluid boundary conditions on body in correspondence with calculated body state and acceleration.\label{stp:updateFluidBCs}
\STATE Update fluid state, ($\vek u, p$), using the fluid solver.
\STATE If end time reached, stop, otherwise go to Step 2.
\end{algorithmic}
\end{algorithm}
In the fluid initialisation in Step \ref{stp:sysInit} and in Step \ref{stp:updateFluidBCs}, it is vital to ensure that proper boundary conditions are specified for the velocity and pressure fields on the body boundary since these contain the coupling between body and fluid (will be detailed in Sections \ref{sec:fluidBCs} and \ref{sec:updateFluidBCs}).

\begin{figure}[tb]
    \centering
     \begin{subfigure}[b]{0.5\textwidth}
         \centering
	    \includegraphics[width=\textwidth]{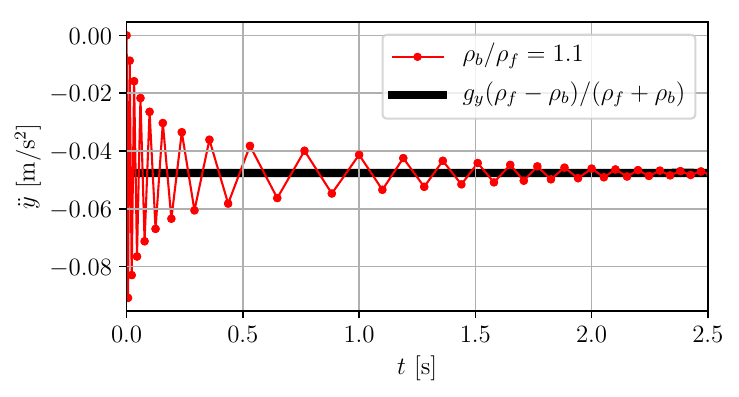}
	    \caption{Acceleration for heavy body.}
	\end{subfigure}
     \begin{subfigure}[b]{0.48\textwidth}
        \centering
	    \includegraphics[width=\textwidth]{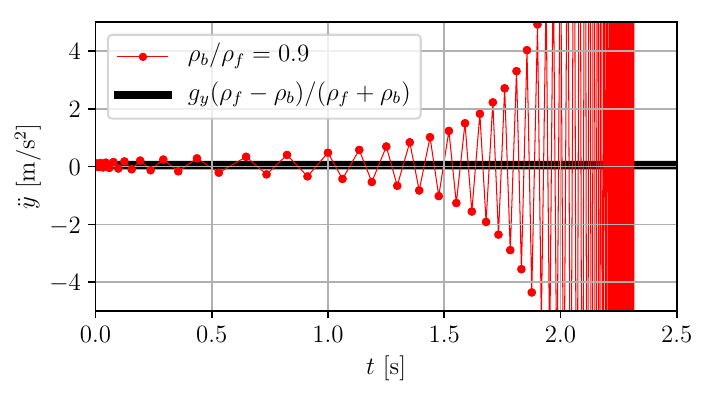}
	    \caption{Acceleration for light body.}
	\end{subfigure}
    \caption{Convergence/divergence of body acceleration in simulation of a circular body of radius $R = 1$ m in gravity $g_y = -1$ m/s$^2$, with density higher/lower than the surrounding inviscid, incompressible fluid ($\rho_f = 1$ kg/m$^3$). Black line marks theoretical value. Outer domain boundary with slip placed at 40$R$ in both cases.\label{fig:buoyantCircleStability}}
\end{figure}

In Step \ref{stp:calca}, we know -- but for now ignore -- that part of the force experienced by the body is due to its instantaneous acceleration, cf. \eqn{eq:Fdivision}. Even if we did not know the specific values of $F_\text{other}$ and $m_a$, we can still explore the implications of ignoring the added mass force in Step \ref{stp:calca}. Using \eqn{eq:Fdivision}, we have
\begin{equation}\label{eq:looseCouplingAccel}
	a_{n+1} = \frac{F_\text{other}}{m_b} - \frac{m_a}{m_b}a_n,
\end{equation}
where the subscript $n$ indicates time step.
If we call $F_\textrm{other}/m_b = a_0$, insert the corresponding expression for $a_n$ in terms of $a_{n-1}$, and so forth until we reach $n = 0$, we get
\begin{equation}
	a_{n+1} = a_0\sum_{k=0}^n\left(-\frac{m_a}{m_b}\right)^k \rightarrow
\begin{cases}
\frac{F_\textrm{other}}{m_b + m_a} & \text{if } m_a < m_b, \\
\pm \infty & \text{if } m_a > m_b, \\
\end{cases}
\end{equation}
i.e. an alternating geometric series bound to diverge in an oscillating manner when the added mass exceeds the body mass. This is the added mass instability in a nutshell. It is an inherent problem in any partitioned coupling mechanism \cite{forsterArtificialAddedMass2007}, and many codes exhibit the instability. This also includes the most widely used open source CFD code, OpenFOAM \cite{devolderReviewImplicitMotion, devolderAcceleratedNumericalSimulations2019, dunbarDevelopmentValidationTightly2015}, which we use in this work as our implementation platform. \fig{fig:buoyantCircleStability} illustrates the change from stable to unstable solver behaviour when the body becomes lighter than the surrounding fluid in the case of a circular body accelerating in a fluid due to gravity. The simulations were run with OpenFOAM's \texttt{inter\-Foam} solver for the fluid motion \cite{deshpandeEvaluatingPerformanceTwophase2012} and the \texttt{six\-DoF\-Rigid\-Body\-Motion} module \cite{huangReviewModellingWaveStructure2022} for body motion.

To circumvent the added mass instability, many codes including OpenFOAM, introduce an outer corrector loop for stronger coupling between body and fluid solution within each time step. In these new iterations, an under-relaxed acceleration is used as shown in \alg{alg:underrelax}.
\begin{algorithm}[h!]
\caption{Strong coupling algorithm with outer corrections and under--relaxation.\label{alg:underrelax}}
\begin{algorithmic}[1]
\STATE Set number of outer correctors, $N_\textrm{OC}$, acceleration relaxation, $\gamma\in [0,1]$ and initial body acceleration, $\vek a$.
\STATE Initialise body state, $(\vek x_b, \vek v_b)$, and fluid state, $(\vek u, p)$.
\STATE Increment time by $\Delta t$\label{stp:+dt}
\STATE Calculate the net force, $\vek F$, on the body including $p$ integrated over its surface
\STATE Store the body acceleration from previous iteration, $\vek a^\textrm{prev} = \vek a$.
\STATE Calculate body acceleration using Newton’s 2nd law: $\vek a^* = \vek F/m_b$
\STATE Under--relax body acceleration, $\vek a = \gamma\vek a^* + (1-\gamma)\vek a^\textrm{prev}$
\STATE Numerically integrate $\vek a$ to get the updated body velocity $\vek v_b^\text{new}$, and position $\vek x_b^\text{new}$.
\STATE Update body state, $(\vek x_b, \vek v_b) = (\vek x_b^\text{new}, \vek v_b^\text{new})$ and mesh. 
\STATE Update fluid boundary conditions on body in correspondence with calculated body state and acceleration.
\STATE Update fluid state, ($\vek u, p$), using the fluid solver.
\STATE If Steps 4-11 were performed less than $N_\textrm{OC}$ times, go to Step 4, otherwise continue.
\STATE If end time reached, stop, otherwise go to Step \ref{stp:+dt}.
\end{algorithmic}
\end{algorithm}

\begin{figure}[tb]
	\centering
     \begin{subfigure}[b]{0.61\textwidth}
         \centering
	    \includegraphics[width=\textwidth]{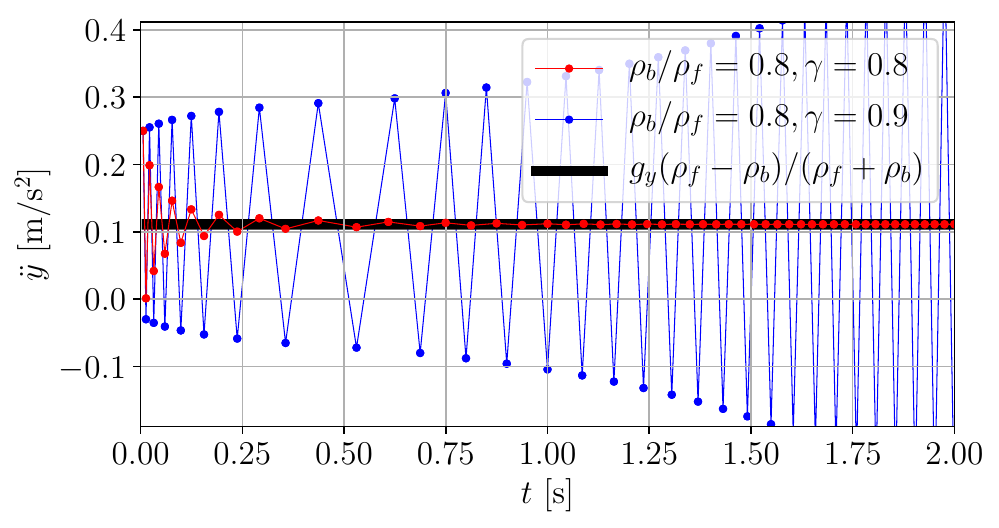}
    \caption{Acceleration as function of time.}
    \label{fig:aRelaxExamples}
     \end{subfigure}
     \begin{subfigure}[b]{0.33\textwidth}
    \includegraphics[width=\textwidth]{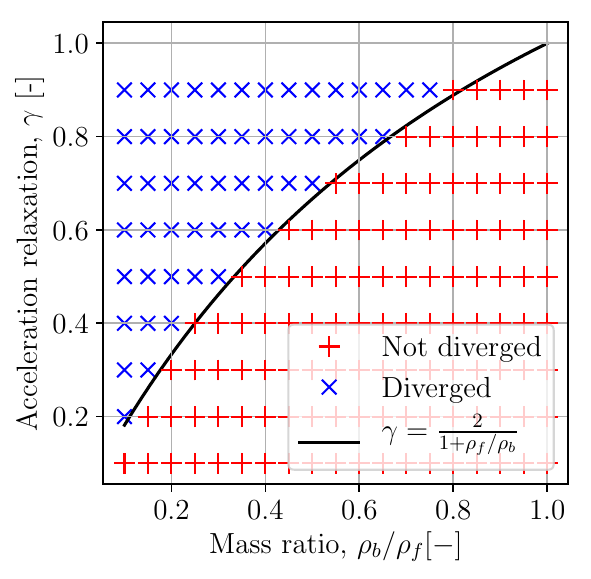}
         \centering
    \caption{Solver stability scan.}
    \label{fig:aRelaxRhoScan}
     \end{subfigure}
    \caption{(a): Circle with $\rho_b = 0.8$ kg/m$^3$ in fluid with $\rho_f = 1$ kg/m$^3$ in gravity $g_y = -1$ m/s$^2$.  Illustration of transition from convergence to divergence when acceleration relaxation, $\gamma$, is changed from $0.8$ to $0.9$ with $\gamma_c \approx 0.0889$. (b): 171 simulations with $\rho_b/\rho_f = [0.1:0.05:1]$ and $\gamma = [0.1:0.1:0.9]$. Blue dot indicates simulation crash due to divergence. Red dot means simulation reached $t = 3 s$ without crash.  Black curve is the theoretical limit from \eqn{eq:gammaCriterion}.\label{fig:aRelax}}
\end{figure}

The introduction of the acceleration relaxation factor, $\gamma\in[0, 1]$, modifies the iterative process in \eqn{eq:looseCouplingAccel} to
\begin{equation}\label{eq:modifiedSeries}
	a_{n+1} = \gamma\left(a_0 - \frac{m_a}{m_b}a_n\right) + (1-\gamma) a_n,
\end{equation}
where the subscript is now an iteration counter rather than a time step counter. Tracking this iterative equation back to the zero'th iteration, we get the modified geometric series,
\begin{equation}
	a_n = \gamma a_0\sum_{k=0}^n\left[1 - \gamma \left(1+\frac{m_a}{m_b}\right)\right]^k.
\end{equation}
This converges if the square bracket has absolute value smaller than one, leading to the stability criterion,
\begin{equation}\label{eq:gammaCriterion}
	\gamma < \gamma_c = \frac{2}{1 + m_a/m_b},
\end{equation}
depending on the instantaneous ratio between added and body mass. This stability criterion was also stated in \cite{soding2001integrate} and derived in a slightly different manner in \cite{devolderAcceleratedNumericalSimulations2019}.
It is clear that the under--relaxation procedure enables convergence for body mass smaller than the added mass. \fig{fig:aRelaxExamples} demonstrates the change to convergence for $\gamma < \gamma_c$ in an OpenFOAM simulation of a light circular body rising in a heavy inviscid fluid. 
\fig{fig:aRelaxRhoScan} shows the stability criterion in \eqn{eq:gammaCriterion} plotted together with numerically obtained convergence tests using OpenFOAM for a scan of acceleration relaxation values and density ratios. This clearly demonstrates that the theoretically obtained convergence criterion indeed applies to the specific body--fluid coupling implemented in OpenFOAM.

When the body--to--fluid density ratio approaches zero, so does $\gamma_c$, and hence also the change in acceleration between iterations diminishes, cf. \eqn{eq:modifiedSeries}. In other words, a larger number of iterations is necessary for light bodies. This becomes computationally expensive, since each outer iteration involves a full CFD update of the fluid state. Also, for problems involving a body near a free surface, or body motion close to other boundaries, the added mass will vary in time, and so will $\gamma_c$ therefore. There exist methods for dynamic relaxation which have been used with some success \cite{dunbarDevelopmentValidationTightly2015, chowStronglyCoupledPartitioned2016}. Here we take a different approach: Instead of trying to fix the trial--and--error approach to finding a consistent body acceleration, we attempt to exploit the freedom CFD grants us to directly measure the zero-acceleration force, $F_\textrm{other}$, in a virtual time step before the actual time step is taken. Likewise, we use a simplified version of the CFD solver, to obtain the instantaneous added mass. This allows us to calculate the body acceleration directly, which is then used to find the new body velocity and position. In this way, we obtain a coupling mechanism that is completely freed from the added mass instability and outer iterations.

\section{Governing equations}
\subsection{Fluid motion}\label{sec:fluidEOMs}
The fluid motion is assumed to be governed by the incompressible Navier-Stokes equations. It may be a single phase with constant density, $\rho_f$, or two immiscible phases separated by a sharp fluid interface across which the fluid mass density jumps from the value $\rho^+$ in the reference fluid to $\rho^-$ in the other fluid. The fluid equations of motion are then
\begin{subequations}\label{eq:fluidEOMS}
\begin{align}
	\frac{\partial \rho}{\partial t} + \nabla\cdot(\rho\vek u) & = 0\label{eq:density}\\
	\nabla\cdot\vek u & = 0\\
	\frac{\partial \rho\vek u}{\partial t} + \nabla\cdot(\rho\vek u\vek u) & = -\nabla p +\rho \vek g + \nabla \cdot(\mu\nabla \vek u) + \vek f,\label{eq:NS}
\end{align}
\end{subequations}
where $\vek g$ is the gravity vector and $\vek f$ represents other forces such as surface tension or external forces. The fluid state at any time is represented by the density field, $\rho$, the pressure field, $p$, and the three components of the velocity field, $\vek u$. As $\rho$, the dynamic viscosity, $\mu$, takes constant values, $\mu^+$ and $\mu^-$ in the the two fluids. Both $\rho$ and the dynamic viscosity, $\mu$, may be expressed in terms of an indicator function, $H(\vek x,t)$, which is a Heaviside function taking the value $1$ in the reference fluid and $0$ in the other:
\begin{equation}
	\rho = \rho^+ H + \rho^- (1-H)\quad \textrm{and}\quad  \mu = \mu^+ H + \mu^- (1-H),
\end{equation}
Eqn.~\eqref{eq:density} can then be replaced by the equivalent equation  
\begin{equation}\label{eq:indicator}
	\frac{\partial H}{\partial t} + \nabla\cdot(H\vek u) = 0.
\end{equation}
This is the starting point for derivations of VoF schemes used to track the sharp fluid interface. For incompressible single--phase flows $\rho$ is constant in the whole domain, and \eqn{eq:indicator} is trivially satisfied.

Above we have used $(\vek u, p)$ to specify the state of the fluid. This is sufficient for single phase flows. For two-phase flows with a sharp fluid interface, we need to augment $\vek u$ and $p$ with a description of the instantaneous fluid interface position. Since this is encoded in the density field, jumping from one value to another at the interface, we will henceforth use the triplet $(\rho, \vek u, p)$ to represent the fluid state.
\subsubsection{Boundary conditions}\label{sec:fluidBCs}
The gradient of the indicator field, $H$, is a 3-dimensional Dirac $\delta$-function which is zero everywhere except on the fluid interface, where it is infinite and points along the interface normal, $\hvek n_I$, into the reference fluid,
\begin{equation}
	\nabla H = \hvek n_I\delta (\vek x - \vek x_I).
\end{equation}
On domain boundaries one can therefore specify a desired contact angle, or rather interface orientation, by specifying $\nabla H$. Often this angle is not of practical interest and one simply uses a zero gradient Neumann boundary condition, $\hvek n_b \cdot \nabla H = 0$, on walls and outlets, where $\hvek n_b$ is the unit normal of the boundary. For inlet boundaries, one must use a Dirichlet boundary condition to specify the interface position of the inflowing fluid.

For the velocity field, $\vek u$, we either use a slip condition or a no-slip condition on walls. For slip we must have $\hvek n_b\cdot\vek u = \hvek n_b\cdot\vek v_b$, where $\vek v_b$ is the velocity of the boundary point. The tangential velocity component can be written $\vek u_t = (I_3 - \hvek n_b\hvek n_b)\vek u$, where $\hvek n_b\hvek n_b$ is the outer product of the vector $\hvek n_b$ with itself. For slip a Neumann condition can be applied for the tangential component, $\hvek n_b\cdot\nabla \vek u_t = 0$. In case of no-slip, the full velocity on the boundary must follow the velocity of the boundary point, $\vek u = \vek v_b$. Again, for inlet boundaries we must specify the velocity value and for domain outlets we can use $\hvek n_b\cdot \nabla \vek u = \vek 0$.

For the pressure, boundary conditions are formulated by requiring consistency with the Navier-Stokes equations \eqn{eq:NS} also on the boundaries of fixed and moving walls. Isolating the pressure gradient and dotting with the boundary normal, we get the Neumann boundary condition,
\begin{equation}\label{eq:pressureBC}
	\hvek n_b \cdot \nabla p  = -\hvek n_b \cdot \left(\frac{\partial \rho\vek u}{\partial t} + \nabla\cdot(\rho\vek u\vek u) - \rho \vek g - \nabla \cdot(\mu\nabla \vek u) - \vek f\right).
\end{equation}
As we will see below, if the boundary is accelerating, care must be taken to encode the acceleration correctly in the implementation of the boundary condition.

\subsection{Rigid body motion}

For the body equations of motion, we start by defining a body-fixed coordinate system centred at some chosen point, $\vek x_0(t)$, and with coordinate axes spanned by the three mutually orthogonal unit vectors $\vek q_1(t), \vek q_2(t)$ and $\vek q_3(t)$. Together, $\vek x_0$ and the orthogonal orientation matrix, $Q = [\vek q_1 \ \vek q_2 \ \vek q_3]$, define the instantaneous configuration of the body. For convenience, we will sometimes denote the body configuration with the shorthand notation,
\begin{equation}
	x_b = \left[\vek x_0\ Q\right].
\end{equation}
The body translational velocity is 
\begin{equation}\label{eq:x0dot}
\dot{\vek x}_0 = \vek v_0(t),
\end{equation}
and the rotational velocity is given in laboratory coordinates by $\bs \omega(t) = [\omega_1\ \omega_2\ \omega_3]^T$ such that
\begin{equation}\label{eq:qdot}
	\dot{\vek q}_i = \bs \omega\times \vek q_i \quad \textrm{for} \quad i = 1,2,3.
\end{equation}
If we define for any vector $\vek v = [v_1\ v_2\ v_3]^T$ the skew--symmetric matrix,
\begin{equation}\label{eq:vtimes}
	\vek v\times = \begin{bmatrix}
0 & -v_3 & v_2\\
v_3 & 0 & -v_1\\
-v_2 & v_1 & 0
\end{bmatrix},
\end{equation}
then \eqn{eq:qdot} can also be written in matrix form as
\begin{equation}\label{eq:Qdot}
	\dot{Q} = \bs\omega\times  Q.
\end{equation}
The acceleration equations are given by Newton's 2nd law
\begin{equation}\label{eq:Newton2nd}
	\frac{d\vek p}{dt} = \vek F, \quad \textrm{and} \quad \frac{d\vek L}{dt} = \bs \tau,
\end{equation}
where the body linear and angular momentum, force and torque (with respect to the laboratory frame origin) are, respectively,
\begin{equation}\label{eq:plFtau}
	\vek p = \int_{\mathcal B}\rho \vek v dV, \quad
	\vek L = \int_{\mathcal B} \vek x\times \rho\vek v dV,\quad 
	\vek F = \int_{\mathcal B}\vek f_\textrm{ext}dV, \quad \textrm{and} \quad 
	\bs\tau = \int_{\mathcal B} \vek x\times \vek f_\textrm{ext} dV.
\end{equation}
Here $\mathcal B$ denotes the body region, and $\vek f_\textrm{ext}(\vek x,t)$ denotes the external force on the body. The velocity in (and on the surface of) $\mathcal B$ is given by
\begin{equation}\label{eq:rigidBodyVelocity}
	\vek v = \vek v_0 + \bs\omega \times (\vek x-\vek x_0).
\end{equation}
We define a body-fixed coordinate system such that the representation of a point in the laboratory frame, $\vek x$, and in the body-fixed coordinates, $\tvek x$, are related by
\begin{equation}
	\tvek x = Q^T(\vek x-\vek x_0) \quad \Leftrightarrow \quad \vek x = \vek x_0 + Q\tvek x.
\end{equation}
Here we have exploited that $Q^{-1} = Q^T$ for the orthogonal matrix, $Q$. Since the body is rigid, the mass density, $\rho$, in $\mathcal B$ is constant in time as viewed from the body-fixed coordinates, i.e. $\rho = \rho(\tvek x)$. The body volume, mass, centre of mass and moment of inertia are, respectively,
\begin{equation}\label{eq:bodyMassDef}
	V_b = \int_{\mathcal B} dV,\quad m_b = \int_{\mathcal B} \rho dV,\quad \vek x_\text{cm}(t) = \frac1{V_b}\int_{\mathcal B}\rho\vek x dV,\quad \tilde I_0 = \int_{\mathcal B} \rho(\tvek x)(|\tvek x|^2 I_3 - \tvek x \tvek x)dV.
\end{equation}
Here $\tilde I_0$ is the moment of inertia with respect to $\vek x_0$, represented in the body--fixed coordinates, and $I_3$ is the 3$\times$3 identity matrix.

The equations for linear and angular acceleration are obtained by inserting Eqns.~\eqref{eq:plFtau} and Eqns.~\eqref{eq:bodyMassDef} in Eqns.~\eqref{eq:Newton2nd} giving
\begin{equation}\label{eq:bodyEOMs}
	\left[\begin{array}{cc}
	m_b I_3 & -m_b Q\tilde{\vek x}_{cm} \times\\
	m_b Q\tilde{\vek x}_{cm}\times  & I_0
	\end{array}\right]
	\left[\begin{array}{c}
	\dot{\vek v}_0\\
	\dot{\bs\omega}
	\end{array}\right]
	= 
	\left[\begin{array}{c}
	\vek F + \bs\omega\times  m_b Q\tilde{\vek x}_{cm}\times \bs\omega\\
	\bs \tau_0 - \bs \omega \times I_0 \bs\omega
	\end{array}\right].
\end{equation}
Note the zero subscript on the torque, $\bs \tau_0$, indicating that it is the torque on the body with respect to the point $\vek x_0$. Together with Eqns.~\eqref{eq:x0dot} and \eqref{eq:Qdot}, \eqn{eq:bodyEOMs} comprise the rigid body equations of motion. For convenience, we will use the more compact notation
\begin{equation}\label{eq:compactNotation}
	M = \left[\begin{array}{cc}
	m_b I_3 & -m_b Q\tilde{\vek x}_{cm} \times\\
	m_b Q\tilde{\vek x}_{cm}\times  & I_0
	\end{array}\right], \quad
	v_b = \left[\begin{array}{c}
	\vek v_0\\
	\bs\omega
	\end{array}\right], \quad
	f = 
	\left[\begin{array}{c}
	\vek F + \bs\omega\times  m_b Q\tilde{\vek x}_{cm}\times \bs\omega\\
	\bs \tau_0 - \bs \omega \times I_0 \bs\omega
	\end{array}\right].
\end{equation}
so that \eqn{eq:bodyEOMs} can be written simply as
\begin{equation}\label{eq:bodyEOMsCompact}
	M\dot v_b = f.
\end{equation}
\subsection{Added mass}
The hydrodynamic force and torque on the body are given by
\begin{equation}\label{eq:hydroForce}
	\vek F_h = \int_{\mathcal S} (-p\vek I_3 + \mu\nabla\vek u)\cdot d\mathbf S,\quad
	\bs \tau_{h,0} = \int_{\mathcal S} (\vek x-\vek x_0)\times (-p\vek I_3 + \mu\nabla\vek u)\cdot d\mathbf S,
\end{equation}
where $\mathcal S =\partial\mathcal B$ is the body surface and $d\vek S$ is the differential area vector pointing out of the body region. Because the fluid is incompressible and the body is rigid, the absolute value of the pressure is immaterial. Only variations in pressure along the surface are relevant for the dynamics.

The pressure parts of the hydrodynamic force and torque contain components, that are proportional to the instantaneous body acceleration, and which go into the total force-torque vector, $f$, in \eqn{eq:bodyEOMsCompact}. Just as we did in the introductory 1-DoF example, we can conceptually split $f$ into a part, $-A\dot v_b$, containing all terms that are proportional to $\dot v_b$, and another part, $f_\textrm{other}$, containing all other forces and torques,
\begin{equation}\label{eq:compactForceEqn}
	f = f_\textrm{other} - A \dot v_b,
\end{equation}
where $A$ is the 6--by--6 added mass matrix, and $f_\textrm{other} \equiv f + A \dot v_b$.

For the added mass matrix, we recall its definition from potential flow theory \cite{milne-thomsonTheoreticalHydrodynamics2011}. In the case of a body moving in an unbounded, incompressible and inviscid fluid, the velocity field can be expressed in terms of a velocity potential, $\vek u = \nabla \phi$. The velocity potential, $\phi$, can be decomposed into contributions proportional to the six velocity components (linear and angular) of the body:
\begin{equation}\label{eq:unitpots}
	\phi = v_1\phi_1 + v_2\phi_2 + v_3\phi_3 + \omega_1\phi_4 + \omega_2\phi_5 + \omega_3\phi_6.
\end{equation}
The functions, $\phi_1,...,\phi_6$ are called the unit potentials because they correspond to unit motion along each of the six degrees of freedom. $\phi_1$ is found by solving a Laplace equation, $\nabla^2\phi_1 = 0$, requiring that $\nabla \phi_1$ approaches zero at infinity and that $\hvek n_b \cdot \nabla\phi_1 = \hvek n_b \cdot (1\ 0\ 0)^T $ m/s on the body boundary. The other unit potentials are found in a similar manner, setting the corresponding body velocity component to one and all other to zero. The added mass matrix can then be expressed in terms of the unit potentials as
\begin{equation}
	A_{ij} = -\int_{\mathcal S}\rho_f \phi_i (\nabla \phi_j)\cdot d\vek S, \quad i,j = 1,...,6.
\end{equation}
In words, the 1st column of the added mass matrix is the linear and angular momentum (or impulse) of the fluid associated with the body moving with unit velocity along the 1st axis in the coordinate system in which the matrix is represented. Likewise, the 2nd and 3rd column contain, respectively, the fluid linear and angular momentum associated with unit body motion along the 2nd and 3rd coordinate axis. The 4th, 5th and 6th columns are populated with the fluid linear and angular momenta corresponding to unit angular velocity around the three coordinate axes, respectively. 

For a body moving in an unbounded fluid the added mass matrix relative to body-fixed coordinates is constant and entirely determined by the body shape. There are many practical situations where the boundaries are so far away that the domain can be regarded as unbounded. 

Even in the presence of viscosity and vorticity, the velocity field can be Helmholtz decomposed into purely potential part and a purely vortical part \cite{eldredgeMathematicalModelingUnsteady2019}, and the added mass force on the body can be shown to be unaltered from the potential flow version \cite{howeFORCEMOMENTBODY1995, concaAddedMassDamping1997}. In other words, the forces and torques on the body from vortices in the fluid -- including vorticity in the boundary layer -- do not depend on the instantaneous acceleration of the body and hence do not contribute to its added mass coefficients. The independence of the added mass on wake vorticity and on the magnitude of the body acceleration has been numerically verified in \cite{wakabaAddedMassForce2007}.

If the domain is bounded, or there are other objects in the fluid, the unit velocity potentials associated with unit linear and angular motion of the body will still be well--defined, but will now depend on the instantaneous geometry of the domain. Hence, as the body moves and reorients relative to the other fluid domain boundaries, the unit potentials and added mass coefficients will also change. Especially if the body pierces a water surface, the instantaneous shape of this surface and the body position and orientation relative to it will influence the added mass matrix. As extreme example is an object falling from air into water, which will give rise to an increase in added mass by a factor of $\rho_\textrm{water}/\rho_\textrm{air} \approx 830$ as the object penetrates the water surface.

\section{The FloatStepper algorithm}

Performing CFD simulations with fixed boundaries, or boundaries moving in a prescribed way, is a standard task performed every day by thousands of engineers and scientists around the world. It is peculiar that adding just six degrees of freedom to the often millions of degrees of freedom used to represent the fluid state can cause severe numerical difficulties. Of all the infinitely many body paths we could prescribe, exactly one corresponds to the path the body would follow if it was free to move in response to the net forces and torques exerted on it, including hydrodynamic forces and other external forces. It is the job of our coupling algorithm to predict the body acceleration that leads us down this particular path when we use it in our prescription of the body motion. What makes this job so hard is the added mass force and its proportionality to the instantaneous body acceleration with a proportionality constant that we do not know in advance. The most widely used methods for solving the implicit acceleration equation is to employ expensive iterations between fluid and body state solvers. Here, we attempt instead to calculate the acceleration directly and non-iteratively. Inserting the decomposed force from \eqn{eq:compactForceEqn} into the body equations of motion, \eqn{eq:bodyEOMsCompact}, and isolating the acceleration, we get
\begin{equation}\label{eq:compactAccelEqn}
	\dot v_b = (M + A)^{-1}f_\textrm{other}.
\end{equation}
This equation is well established in linear radiation-diffraction theory for floating bodies \cite{newmanMarineHydrodynamics}, where the added mass can be computed at the bodies equilibrium position. For it to be useful in unsteady CFD, and without assumption of small waves or body motion, we need to devise methods to calculate the unknown and time dependent vector, $f_\textrm{other}$, and matrix $A$.
\subsection{The zero--acceleration step}
Let us think of a snapshot of our body-fluid system with body state $(x_b,v_b)$, fluid state, $(\rho, \vek u, p)$, and possibly a number of external forces acting on the body. At this point in time, we need to decide where the body should go in the next time step in order for its motion to represent free motion. Suppose we took a time step with the same $v_b$ as in the previous time step. This would be experienced by the fluid as a step with zero acceleration, $\dot v_b = 0$. According to \eqn{eq:compactForceEqn}, the force experienced by the body during such a zero--acceleration time step would be
\begin{equation}\label{eq:fzero}
	f = f_\textrm{other}.
\end{equation}
In other words, taking a zero--acceleration time step with our CFD solver, and measuring the resulting hydrodynamic response force and torque, reveals the non-added mass part of \eqn{eq:hydroForce}, which, together with gravity, mooring lines, self propulsion etc., comprises $f_\textrm{other}$. 
\subsection{Rewinding system}
In our process of developing FloatStepper, we initially attempted to take the zero--acceleration time step without actually moving the mesh, as we otherwise do in real CFD time steps. This was, however, found to lead to wrong estimates of $f_\textrm{other}$. Instead we take the zero--acceleration time step using mesh motion and exactly the same CFD solver settings as in the real time step. This ensures accurate estimation of $f_\textrm{other}$. It also requires a careful time reversal step where, once $f_\textrm{other}$ is obtained, the fluid, body and mesh are brought back exactly to their state before the zero--acceleration time step.
\subsection{Added mass estimation}
To numerically measure the instantaneous added mass matrix, we exploit its definition as the constant of proportionality between hydrodynamic force and body acceleration. We also exploit that the added mass in a viscous fluid with vorticity is identical to the one obtained in the corresponding potential flow situation. In the added mass calculation the convective and viscous terms can be neglected (see e.g. \cite{mouginGeneralizedKirchhoffEquations2002}), so the equation to solve is simply
\begin{equation}\label{eq:addedMassEqn}
	\frac{\partial \rho \vek u}{\partial t} = -\nabla p
\end{equation}
We now discretise the time derivative using the Euler scheme,
\begin{equation}\label{eq:EulerDdt}
	\frac{\partial \rho \vek u}{\partial t} \approx \frac{\rho^{n+1}\vek u^{n+1}-\rho^n\vek u^n}{\Delta t},
\end{equation}
where the superscript denotes time step. In potential flow theory, the added mass associated with motion along the $x$-axis is obtained by impulsively changing the body velocity from zero to $\vek v_0 = (1, 0, 0)$ m/s amounting to a boundary condition, $\vek n_b\cdot \vek u = \vek n_b\cdot (1,0,0)$ m/s for $\vek u^{n+1}$. Inserting \eqn{eq:EulerDdt} in \eqn{eq:addedMassEqn} and taking the divergence, we get
\begin{equation}\label{eq:pAddedMassEqn}
	\nabla\cdot\left(\frac1{\rho^{n+1}}\nabla (\Delta t p_1)\right) = 0,
\end{equation}
where we have used that $\vek u^n = \vek 0$, required $\nabla\cdot\vek u^{n+1} = 0$, and marked the pressure with a subscript $1$ to indicate that it is the pressure corresponding to acceleration $a_1 = 1 $m/s$/\Delta t$ along the first degree of freedom (here chosen to be the $x$-axis). We have also collected $\Delta t p_1$ in a bracket in \eqn{eq:pAddedMassEqn} to emphasise that this impulse approaches a constant as $\Delta t \rightarrow 0$. Physically, this means that if we impose the body velocity over a very short (long) time step, $\Delta t$, then the pressure amplitude required to set the fluid into corresponding motion must be very high (low) such as to keep $\Delta t p$ constant. In the limit $\Delta t\rightarrow 0$, the density $\rho^{n+1}$ approaches $\rho^n$, and so in our added mass calculation step, we do not update the fluid interface position encoded in $\rho$, i.e. we simply use $\rho^n$ instead of $\rho^{n+1}$ in \eqn{eq:pAddedMassEqn}.

Once the pressure, $p_1$, corresponding to the acceleration, $a_1 = 1$ m/s $/\Delta t$ of the body along the $x$-axis is found, the corresponding force and torque on the body are calculated as
\begin{equation}
	f_1 = \int_{\mathcal S} p_1 \begin{bmatrix}I_3 \\ (\vek x -\vek x_0)\times \end{bmatrix}d\vek S.
\end{equation}
and the first column of the added mass matrix is then given by
\begin{equation}
	A_1 = -f_1/a_1.
\end{equation}
The second to sixth column of the added mass matrix are calculated in the same way, calculating the pressure force corresponding to unit body velocity along the other two axes, and unit angular velocity around the three coordinate axes. 

In our current implementation of the added mass matrix calculator, the boundary conditions on all other boundaries than the rigid body are copied from the fields used in the real time step. Solver settings for the pressure equation are also copied from the real time step pressure solution. The added mass calculator is implemented as a copy of the PISO step in the interFoam solver except that the convective and viscous terms have been removed from the momentum equation. Thus, the calculation is fully parallelised, using the same domain decomposition as the real time step. The method allows the user to specify which degrees of freedom should be active in a simulation, for instance first, second and sixth for a body freely moving and rotating body in the $xy$-plane. It also allows the user to specify a parameter, \texttt{Madd\-Update\-Freq}, to only update the added mass matrix every \texttt{Madd\-Update\-Freq}'th time step. This may save computation time in simulations where the added mass is known to only change slowly. Finally, we mention that the added mass calculator class is derived from an abstract base class, allowing future addition of alternative added mass calculator classes (for instance a panel method based calculator) with runtime selection of the preferred method specified in the case setup files.
\subsection{Body state update}
Once $f_\textrm{other}$ and $A$ have been calculated, as described above, the body acceleration is calculated directly from \eqn{eq:compactAccelEqn}. This brings us to the actual integration of the body acceleration and velocity to obtain the new velocity and position. This can be done with standard ODE solvers, and while the choice of integration scheme here can have important consequences for energy conservation etc., it is not a point of focus for our work here. We simply use Euler integration, $x^{n+1} = x^{n} + \dot{x}^n\Delta t$, except for the orientation matrix, $Q$, which we update using the Rodrigues rotation formula,
\begin{equation}
	Q^{n+1} = \left(I_3 + \sin(|\bs\omega|\Delta t)\left(\frac{\bs\omega}{|\bs\omega|} \times\right)  + \left[1-\cos(|\bs\omega|\Delta t)\right]\left(\frac{\bs\omega}{|\bs\omega|} \times\right)^2\right)Q^n,
\end{equation}
to ensure that it stays orthogonal.
\subsection{Mesh motion}
From the fluid side, the body is represented by a boundary patch on which discretised versions of the boundary conditions in Section \ref{sec:fluidBCs} are applied. Thus, after the body position and velocity have been updated to their newly found values, the mesh must follow along. In our current implementation, we use the deforming (or ``morphing'') mesh functionality of OpenFOAM with the body boundary patch moving rigidly, and the mesh points in a region around the patch deforming to accommodate this motion. In this approach all mesh points closer to the body patch than a user defined \texttt{inner\-Distance} follow the body in its rigid body motion. Mesh points outside a user specified \texttt{outer\-Distance} from the body patch are kept stationary. Mesh points between these two distances adapt their position smoothly using Spherical Linear Interpolation (SLERP) based on their distance to the body \cite{shoemakeAnimatingRotationQuaternion1985}. This leads to acceptable mesh quality as long as the body displacement and rotation are not too large.

\subsection{Updating boundary conditions on a rigid body}\label{sec:updateFluidBCs}

In the meshed fluid domain the boundary patch representing the rigid body consists of polygonal faces, each with a face centre, $\vek x_f$, which is updated in each time step. We have implemented a velocity boundary field class called \texttt{floater\-Velocity}, which holds a reference to a floating body object from which it reads a body position, $\vek x_0$, velocity, $\vek v_0$, and angular velocity, $\bs\omega$. For each face on the rigid body boundary patch it then sets the velocity field using \eqn{eq:rigidBodyVelocity} with $\vek x = \vek x_f$. The boundary condition takes a boolean parameter called \texttt{slip}. If this is set to \texttt{true}, the boundary condition only takes the normal component from \eqn{eq:rigidBodyVelocity}. The velocity component tangential to the face is directly copied from the tangential component of the velocity vector at the centre of the cell to which the face belongs. This is to lowest order (ignoring the curvature of the surface) a symmetry condition on the tangential velocity component.

In the rigid body pressure boundary condition, \eqn{eq:pressureBC}, the dependency on body acceleration appears indirectly in the first term on the right hand side. This can be seen by writing it as
\begin{equation}
	\frac{\partial \rho\vek u}{\partial t} 
	= \rho \dot{\vek v} + \vek v\frac{\partial\rho}{\partial t} 
	= \rho\left[\dot{\vek v}_0 + \dot{\bs\omega}(\vek x - \vek x_0) 
	+ \bs\omega\times\{\bs\omega\times(\vek x - \vek x_0)\}\right] 
	+ \left[\vek v_0+\bs\omega\times(\vek x - \vek x_0)\right]\frac{\partial\rho}{\partial t},
\end{equation}
where we have inserted and differentiated the rigid body velocity, $\vek v$, from \eqn{eq:rigidBodyVelocity}. This acceleration dependency is the very origin of the added mass force on the body and therefore is important to capture. We remark that in OpenFOAM, this dependency is treated indirectly by the PISO solution procedure. Discretising the momentum equation, \eqn{eq:NS}, it can be written
\begin{equation}
	a^{\vek u}\vek u^{n+1} = \vek H - \nabla p,
\end{equation}
where $a^{\vek u}\vek u^{n+1}$ is a collection of all terms proportional to the new time velocity, $\vek u^{n+1}$, and $\vek H$ (a vector not to be confused with the indicator field introduced in Section \ref{sec:fluidEOMs}) contains all other terms except the pressure gradient, $\nabla p$. If the time derivative in \eqn{eq:NS} is for instance discretised using the Euler method (\eqn{eq:EulerDdt}), then $a^{\vek u}$ will contain a term, $\rho^{n+1}/\Delta t$, and $\vek H$ will contain a term, $\rho^n\vek u^n/\Delta t$. For boundary faces on the rigid body, these old and new velocities are determined by the specified body acceleration, which is then conveyed to the pressure by imposing a boundary condition for $p$ on the body surface of the form,
\begin{equation}\label{eq:pBC}
	\vek n_b\cdot\nabla p = \vek n_b \cdot \left(\vek H/a^{\vek u} - \vek u^{n+1}\right)a^{\vek u}.
\end{equation}
Here, for the acceleration to be included correctly, care must be taken when building the $\vek H/a^{\vek u}$-field, and imposing boundary conditions on it. In particular, when assembling the pressure Poisson equation, the standard OpenFOAM solver calls a function, \texttt{constrain\-HbyA}, which sets $\vek H/a^{\vek u}$ equal to $\vek u^{n+1}$ on boundaries where the $\vek u$ boundary condition is of Dirichlet type, including rigid bodies. This results in $\vek n_b\cdot \nabla p = 0$, which we have just argued is incorrect when the body is accelerating. In our simulations, we have found this to give rise to an erroneous behaviour, where the velocity in the cell layer closest to the accelerating body has only around half of the correct magnitude regardless of the thickness of this layer. To avoid this, we make use of the build--in pressure boundary condition called \texttt{fixed\-Flux\-Extrapolated\-Pressure}. Using this, $\vek H/a^{\vek u}$ retains its boundary value obtained as a sum of all the terms from which it is composed, including the previous time velocity, $\vek u^n$, and we no longer observe the erroneous velocity behaviour. 

\subsection{Fluid state update}

The exact procedure for solving the fluid equations is not the focus of the work presented here, and will therefore only be described briefly. The implementation is based on OpenFOAM's interfacial flow solver, \texttt{interIsoFoam}, (version v2206) employing the finite volume method to solve the motion of two immiscible fluids on arbitrary unstructured meshes with cell-centred collocated field representation.

The fluid solver for updating the fluid state, $(\rho,\vek u, p)$, starts by updating the fluid interface position using the VoF method, where the interface is represented by a volume fraction field expressing for each cell how much of its volume is occupied by the reference fluid. There exist many VoF methods, and the FloatStepper algorithm does not depend on this choice. Here we use the geometric VoF method called isoAdvector, which ensures a sharp interface and accurate, efficient interface advection on arbitrary unstructured meshes \cite{roenbyComputationalMethodSharp2016, roenbyIsoAdvectorGeometricVOF2019}. The solver can be run in single-phase mode simply by setting the volume fraction field to 1 in all cells and on all boundaries. In this case, isoAdvector will find no interface cells and hence will do nothing.

After the interface advection step the pressure and velocity fields are updated using the PISO algorithm \cite{issaSolutionImplicitlyDiscretised1986}. Details of the OpenFOAM specific solution procedure can be found in \cite{deshpandeEvaluatingPerformanceTwophase2012, moukalledFiniteVolumeMethod2016, uroic2019implicitly}. When the mesh is moving, the convection of mass and momentum in \eqn{eq:fluidEOMS} is made relative to the mesh motion. This is done by subtracting the flux due to motion of mesh faces from the physical fluxes across faces in the discretised convective terms, as described e.g. in \cite{jasak2010dynamic}.

\subsection{Summary of algorithm}

In \alg{alg:FloatStepper1D} we summarize the FloatStepper coupling algorithm. We remark that a similar approach to separating the force into an added mass contribution and everything else was briefly and elegantly sketched by S\"oding in \cite{soding2001integrate}. Our method differs from S\"oding's by being non-iterative. The added mass estimate in S\"oding's algorithm is found via a minimisation process and used as a good initial guess for an iterative solution procedure for the implicit acceleration equation. We, on the other hand, attempt to calculate $f_\textrm{other}$ and $A$ directly and sufficiently accurately, so that iterations can be avoided. S\"oding provides no validation and only few implementation details. In \cite{devolderAcceleratedNumericalSimulations2019} Devolder et al. presented a 1 DoF OpenFOAM implementation of a similar iterative approach, and showed its favourable stability properties for a heaving floater.
\begin{algorithm}[h!]
\caption{The FloatStepper algorithm.\label{alg:FloatStepper1D}}
\begin{algorithmic}[1]
\STATE Initialise the body state, $(x_b, v_b)$, and fluid state, $(\rho,\vek u, p)$.
\STATE Increment time by $\Delta t$ \label{stp:FStimeInc}
\STATE Take a probe time step with zero acceleration and measure the resulting force and torque, which by \eqn{eq:fzero} is equal to $f_\textrm{other}$.\label{stp:zeroAccSteo}
\STATE Rewind body, mesh and fluid states to state just before Step \ref{stp:zeroAccSteo}
\STATE Calculate updated added mass matrix, A (optionally, only every \texttt{MaddFreqUpdate}'th time step).
\STATE Calculate body acceleration as $\dot v_b = (M + A)^{-1}f_\text{other}$.
\STATE Time integrate $\dot v_b$ to get $v_b^\text{new}$, and $v_b$ to get $x_b^\text{new}$.
\STATE Move body (and mesh) accordingly and update fluid boundary conditions on its surface.
\STATE Calculate new fluid state as if the found body displacement was prescribed.
\STATE If end time reached, stop, else go to Step \ref{stp:FStimeInc}.
\end{algorithmic}
\end{algorithm}
\\
\section{Validation}

\subsection{Light disc in gravity}
In the introduction, we illustrated the added mass instability with a light disc rising in a heavy, inviscid fluid. We recall that in such an ideal fluid the force on the body is obtained by integrating the pressure over the body surface with the pressure given by the unsteady Bernoulli equation,
\begin{equation}\label{eq:Bernoulli}
	p = -\rho_f\frac{\partial \phi}{\partial t} -\rho_f \frac12|\vek u|^2 + C,
\end{equation}
where $C$ is an arbitrary constant and $\phi$ is the velocity potential. For pure translational motion along the $y$-axis $\phi$ can be written as $v_y\hat{\phi}_y$, where $\hat{\phi}_y$ is the unit velocity potential associated with unit body velocity along the $y$-axis. The unit potential can be obtained by employing the Milne-Thomson circle theorem \cite{achesonElementaryFluidDynamics1990}.
Thus, it is the time derivative of $\phi$ in \eqn{eq:Bernoulli}, which gives the added mass force contribution  proportional to the instantaneous acceleration, $\dot v_y$, and to $\rho_f$, when integrated over the body surface. Isolating the acceleration in the resulting force expression, one obtains the theoretical constant  body acceleration \cite{milne-thomsonTheoreticalHydrodynamics2011},
\begin{equation}\label{eq:a_y}
	a_y = \frac{\rho_f-\rho_b}{\rho_f+\rho_b}g_y.
\end{equation}
We remark that the $\rho_f$ in the denominator comes from added mass term and prevents the acceleration from diverging when $\rho_b/\rho_f\rightarrow 0$. This term is sometimes omitted in the literature, for instance in section 13.10 of \emph{Computational Methods for Fluid Dynamics} by Ferziger et al. \cite{ferzigerComputationalMethodsFluid2019} although their example has $\rho_b/\rho_f = 0.01$. It is exactly this omission in numerical coupling methods that leads to large overestimation of the body acceleration and numerical instability caused by unphysical kinetic energy injection.
\begin{figure}[tb]
    \centering
     \begin{subfigure}[b]{0.7\textwidth}
         \centering
         \includegraphics[width=\textwidth]{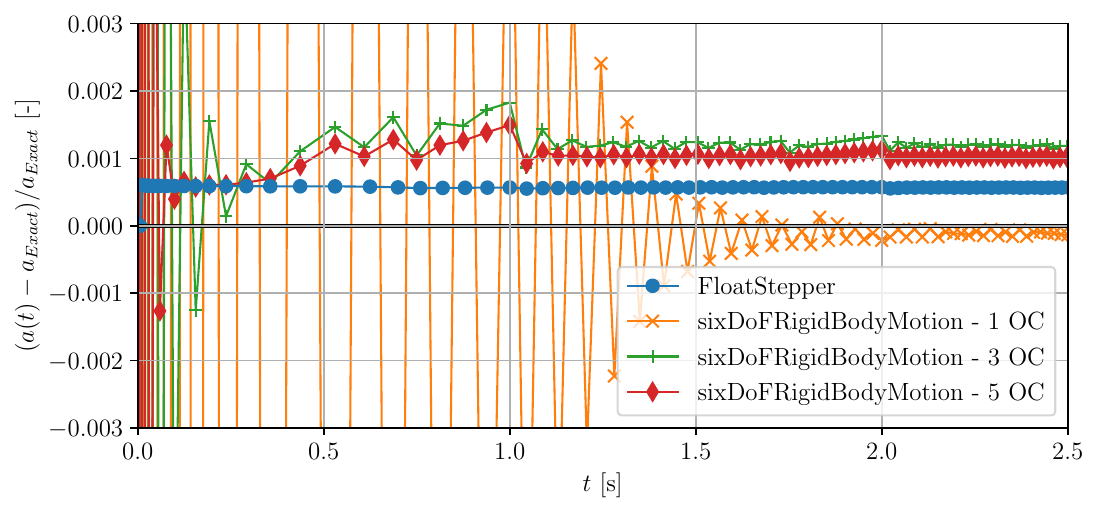} \\
			\caption{\label{fig:fstpVs6oF}}
		\end{subfigure}
     \begin{subfigure}[b]{0.28\textwidth}
         \centering
		   \includegraphics[width=\textwidth]{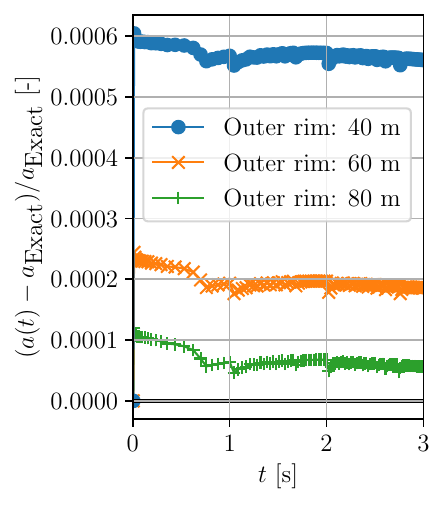}
			\caption{\label{fig:fstpEnlargeDomain}}
		\end{subfigure}
    \caption{(a) Relative error in predicted acceleration for a circle of radius $1$ m and density $\rho_b = 0.8$ kg/m$^3$ surrounded by inviscid fluid of density $\rho_f = 1$ kg/m$^3$. The \texttt{six\-DoF\-Rigid\-Body\-Motion} simulation is done with 1, 3 and 5 outer correctors (the former identical to the stable solution shown in Fig.~\ref{fig:aRelaxExamples}), acceleration relaxation $\gamma = 0.8$ and Euler time integration of the body equations of motion. FloatStepper and \texttt{six\-DoF\-Rigid\-Body\-Motion} simulations use the same mesh and adaptive time step with maximum CFL of 0.5. (b) FloatStepper simulation repeated with outer domain at 40 m (same as in (a)), 60 m and 80 m. \label{fig:buoyantCircleFloatStepper}}
\end{figure}

\fig{fig:fstpVs6oF} shows the relative acceleration error (relative to $a_y$ from \eqn{eq:a_y}) as a function of time for a FloatStepper simulation (blue), where the light circle is released to rise buoyantly at time zero. The FloatStepper acceleration is very close to constant with a relative error of around 0.06 \%. For comparison we also show in \fig{fig:fstpVs6oF} the results obtained with the \texttt{six\-DoF\-Rigid\-Body\-Motion} library of OpenFOAM with 1, 3 and 5 outer correctors in \alg{alg:underrelax}. For those simulations the initial acceleration is miscalculated due to the initial estimate $a = F/m_b$ built into the algorithm. Increasing the number of outer iterations makes the simulation converge faster, but we cannot completely avoid the faulty initial accelerations, and the computational cost increases in proportion to the number of outer correctors. 

The \texttt{six\-DoF\-Rigid\-Body\-Motion} results in \fig{fig:fstpVs6oF} are run with acceleration relaxation $\gamma = 0.8$ based on our knowledge of $\gamma_c$ from \eqn{eq:gammaCriterion} and \fig{fig:aRelaxRhoScan} to ensure convergence. We remind the reader that this was only possible because of the simplicity of the case, a circle with known, constant added mass. In practical simulations, the added mass is normally not known and may vary with time. Indeed, one of the frustrating aspects of working with the \texttt{six\-DoF\-Rigid\-Body\-Motion} is the guesswork going into setting the acceleration relaxation and the number of outer correctors for a given simulation situation. The user often faces a choice between excessive simulation time and reduced accuracy at best, or numerical instability at worst. Eliminating this guesswork is one of the main motivations for developing FloatStepper.

We have numerical investigated the cause of the constant 0.06\% error of FloatStepper in \fig{fig:fstpVs6oF}. We have found that the error is unaltered by reducing the time step size or increasing the mesh resolution. The simulation was done with a circle of radius 1 m and with the circular outer rim of the domain placed 40 radii away. We have found that if we repeat the numerical experiment with radius of 60 m and 80 m instead of 40 m, the observed deviation from the theoretical value is reduced, see \fig{fig:fstpEnlargeDomain}. This suggests that the deviation from the theoretical (infinite domain) value, is in fact not a numerical error but rather a finite domain size effect.

It is a peculiar thought that even at the very moment when the body is released, but has no velocity yet, it experiences a large added mass force from the surrounding fluid due to the momentum the fluid were to gain, if it was set into motion by the body. This is an artifact of the strict fluid incompressibility condition, causing the fluid and body to be as tightly dynamically coupled as the two halves of the rigid body. In reality no physical fluid is perfectly incompressible, so any real body acceleration will cause a pressure wave with corresponding fluid compression to emanate from the body at the speed of sound. Indeed, fluid solver approaches like Smoothed--Particle Hydrodynamics (SPH), where an artificial compressibility is introduced, do not appear to suffer from the added mass instability.

\begin{figure}[h!]
    \centering
	\begin{subfigure}[b]{0.48\textwidth}
    	\centering
	    \includegraphics[width=\textwidth]{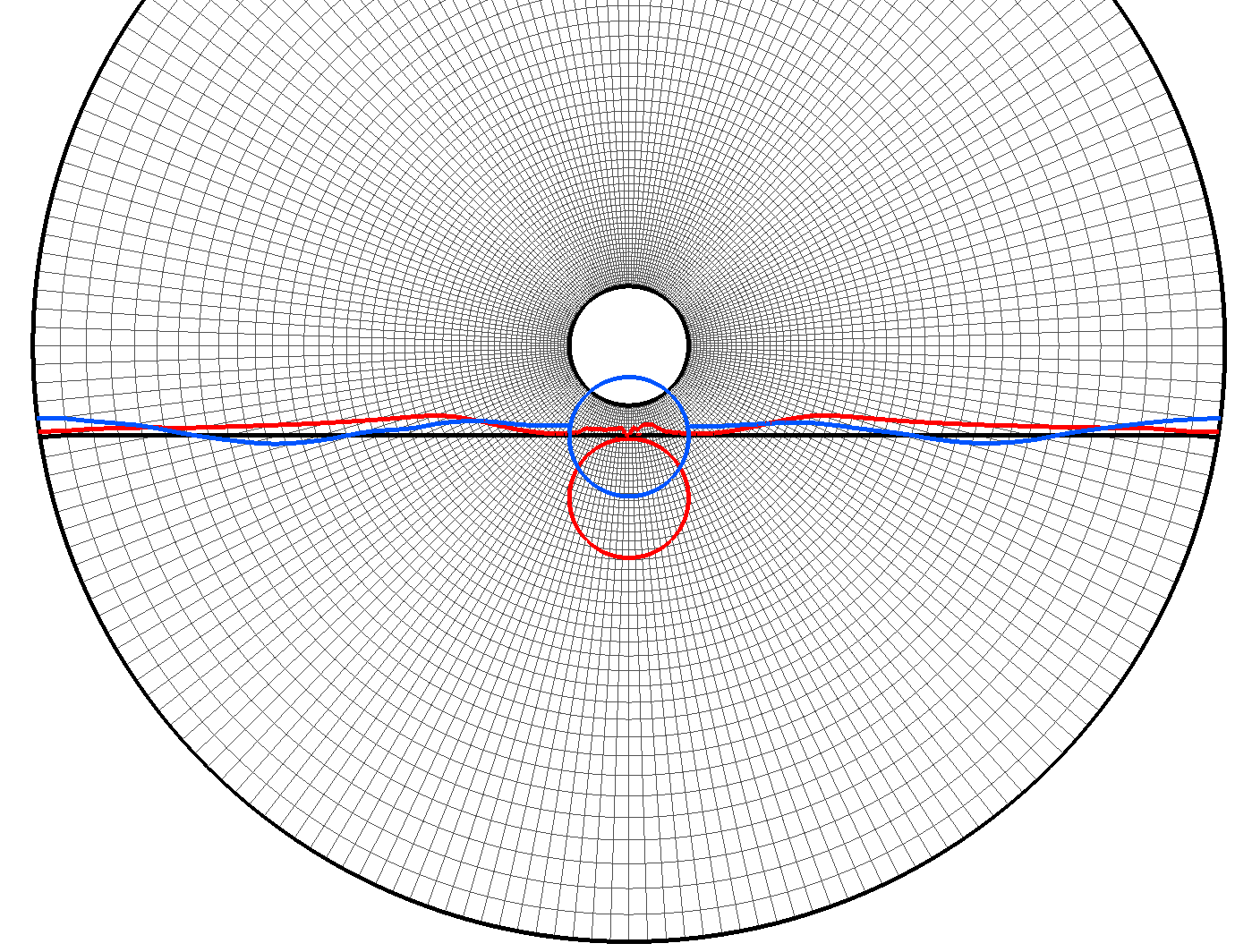}
	    \caption{\label{fig:fallingCircle}}
	\end{subfigure}
	\begin{subfigure}[b]{.50\textwidth}
    	\centering
	    \includegraphics[width=\textwidth]{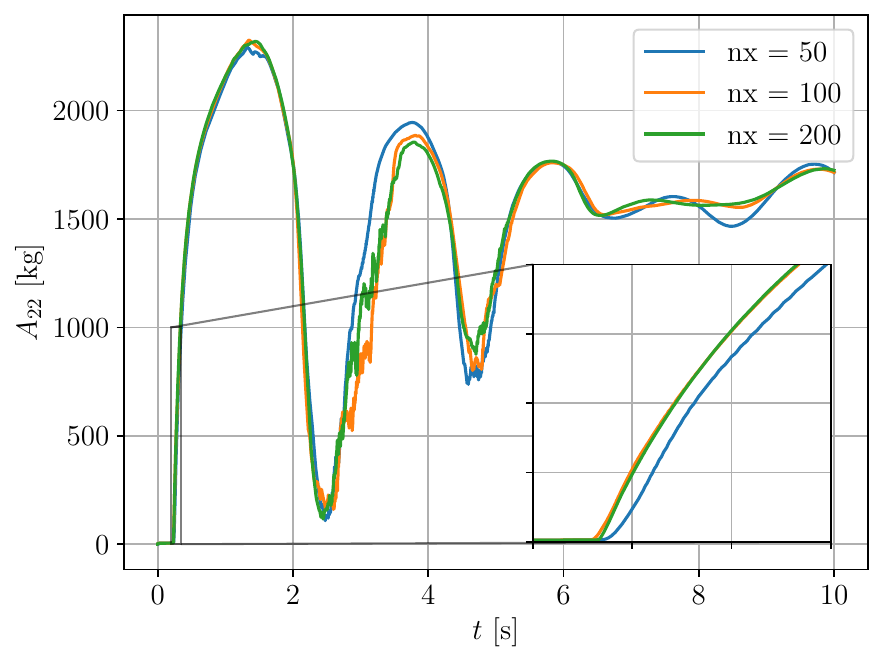}
	    \caption{\label{fig:MaddConv}}
	\end{subfigure}
    \caption{(a) A circular domain of radius 10 m centred at the origin and a water surface placed at $y = -1.5 m$. A circular body of radius $R = 1$ m is initialised at the origin with downward velocity $v_y = -1$ m/s. Gravity is $g_y = -9.81$ m/s$^2$, air density is $\rho_a = 1$ kg/m$^3$, water density is $\rho_w = 1000$ kg/m$^3$, and the body has density $\rho_b = 500$ kg/m$^3$. Body position and water surface shown for time $t = 0$ s (black), $t = 1.3$ s (red) and $t = 10$ s (blue). (b) Evolution of the vertical added mass component with time for three different mesh resolutions.\label{fig:fallingCircle}}
\end{figure}

\subsection{Disc in gravity hitting water surface}

In the previous case we recalculated the added mass at every time step although it was essentially constant due to the long distance to boundaries and absence of a fluid interface. We now test our added mass calculation procedure with a test case involving a large sudden change in added mass, namely a circular body falling from air into water. Only the vertical component of the body motion is active. \fig{fig:fallingCircle} shows the initial configuration (black) as well as the body position and water surface at time $t = 1.3 s$ (red), where the body is fully immersed in water, and at the end of the simulation, $t = 10 s$ (green). In \fig{fig:MaddConv} we show the time evolution of the added mass during the simulation for three different mesh resolutions. As expected, we observe a sudden rise in added mass as the body hits the water surface with a maximum when the body is fully immersed in water. The entry of the body into the water creates waves that are reflected at the domain walls, causing an irregular heaving motion of the body as the added mass settles to its equilibrium value dictated by the density ratios. The convergence with mesh resolution in the initial phase, where the body hits the water surface, is very good as shown in the inset of \fig{fig:fallingCircle}. At later stages the correspondence between the three simulations is also good, although with small variations, presumably due to difference at different mesh resolutions in the ability to capture the details of the complicated, reflected wave field. This example demonstrates how the FloatStepper algorithm is able to robustly handle large and abrupt changes in added mass.

\begin{figure}[h!]
    \centering
    \begin{subfigure}[b]{\textwidth}
    	\centering
	    \includegraphics[width=\textwidth]{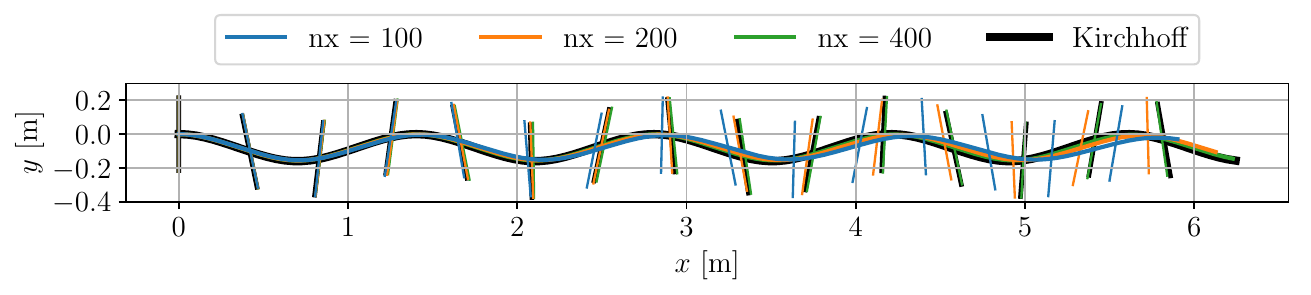}
	    \caption{\label{fig:wigglingTrajs}}
	\end{subfigure}
    \begin{subfigure}[b]{0.61\textwidth}
    	\centering
	    \includegraphics[width=\textwidth]{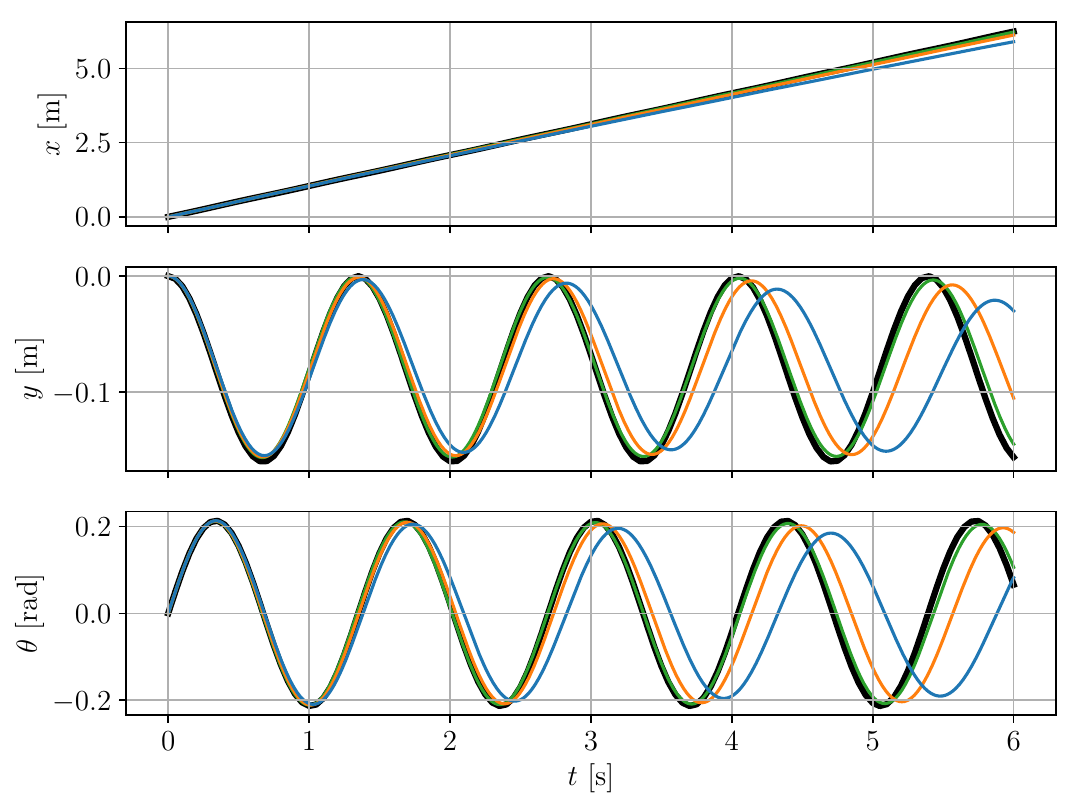}
	    \caption{\label{fig:wigglingCoord}}
	\end{subfigure}
    \begin{subfigure}[b]{0.375\textwidth}
    	\centering
	    \includegraphics[width=\textwidth]{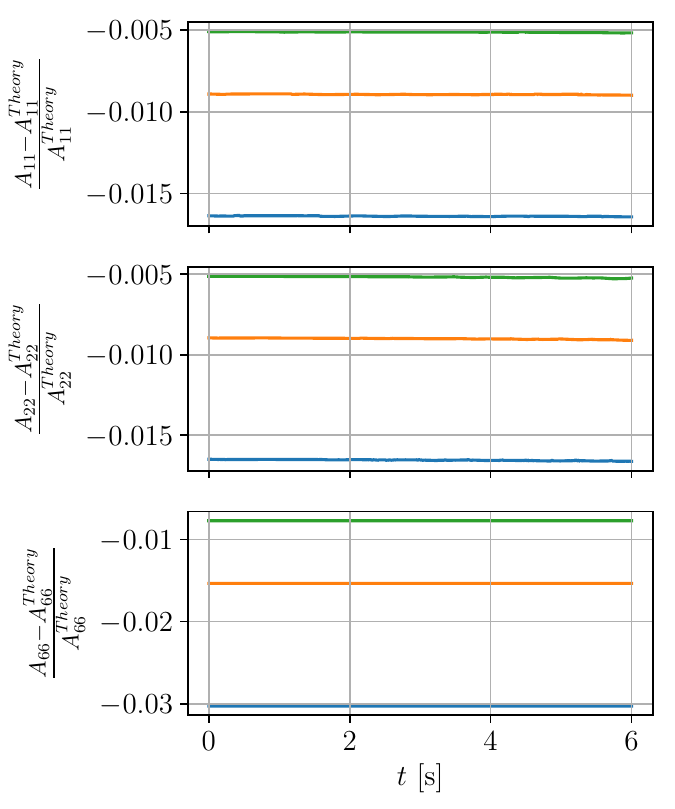}
	    \caption{\label{fig:wigglingMadd}}
	\end{subfigure}
    \caption{(a) Trajectory of an elliptic body of density $\rho_b = 0$ kg/m$^3$ moving through an infinite 2D ideal fluid of density $\rho_f = 1$ kg/m$^3$. Initial body centre at $(0,0)$ with body minor-axis aligned with the $x$-axis. Initial velocity is $(v_x,v_y,\omega)= (1, 0, 1)$. Black curve is exact solution while simulations with three different mesh resolutions are shown in colours. Body orientation (major axis) shown at 15 locations along the trajectory. (b) Body coordinates as function of time. (c) Relative deviation of added mass coefficients from infinite domain values in \eqn{eq:A_ii}.  \label{fig:wigglingEllipseNxScanTraj}}
\end{figure}

\subsection{Free ellipse in infinite ideal fluid}
When a rigid body free to translate and rotate is immersed in a fluid, the hydrodynamic forces introduce a coupling so that translation can induce rotation and vice verse. It is important to verify that our algorithm captures this coupling correctly. To this end, we consider a benchmark case with a rigid body moving through inviscid fluid with all boundaries far away. According to Howe \cite{howeFORCEMOMENTBODY1995} the hydrodynamic force associated with the body motion can be written
\begin{subequations}\label{eq:FM_h}
\begin{align}
	\vek F_h & = -\frac{\partial \ }{\partial t}(T\vek v_0 + S\bs\omega) \\
	\bs\tau_{h,0} & = -\frac{\partial \ }{\partial t}(S^T\vek v_0 + J\bs\omega) - \vek v_0\times(A\vek v_0 + C\bs\omega)
\end{align}
\end{subequations}
where $T, S$ and $J$ are the $3\times 3$ added mass submatrices,
\begin{equation}
A = \begin{bmatrix}
T & S\\
S^T & J
\end{bmatrix}.
\end{equation}
To see how these added mass coefficients change in time, we note that the matrices $T, S$ and $J$ represented in the body-fixed basis, $\{\vek q_1,\vek q_2, \vek q_3\}$, are constants in time determined solely by the body geometry.  We will call these matrices $\hat T, \hat S$ and $\hat J$ and note that we have $T = Q\hat TQ^T$ and so on. Then for instance the first term in \eqn{eq:FM_h} becomes
\begin{equation}
	\partial_t (T\vek v_0) = \dot Q \hat T Q^T\vek v_0 + Q\hat T \dot Q^T \vek v_0 + Q\hat T Q^T\dot{\vek v}_0.
\end{equation}
Noting that $\dot Q = \bs \omega\times Q$ and that $(\bs\omega\times)^T = -\bs\omega\times$, we can rewrite this to
\begin{equation}
	\partial_t (T\vek v_0) = \bs \omega\times T\vek v_0 - T \bs \omega\times \vek v_0 + T \dot{\vek v}_0.
\end{equation}
Using this for all the matrix-vector products in \eqn{eq:FM_h}, we get
\begin{equation}\label{eq:KKforce}
		\begin{bmatrix}
		\vek F_h \\
		\bs\tau_{h,0}
	\end{bmatrix}
	=
	\left(
	A\begin{bmatrix}
		\bs \omega\times & 0_3 \\
		0_3 & \bs \omega\times 
	\end{bmatrix}
-
	\begin{bmatrix}
		\bs \omega\times  & 0_3 \\
		0_3 & \bs \omega\times 
	\end{bmatrix}A
	\right)
	\begin{bmatrix}
		\vek v_0 \\
		\bs\omega
	\end{bmatrix}
-
	A\begin{bmatrix}
		\dot{\vek v}_0 \\
		\Dot{\bs\omega}
	\end{bmatrix}
-  
	\begin{bmatrix}
		\vek 0 \\
		\vek v_0\times (T\vek v_0 + J\bs\omega)
	\end{bmatrix}.
\end{equation}
Inserting this as the force and torque in the body equations of motion, \eqn{eq:bodyEOMs}, one obtains the Kirchhoff equations \cite{lambHydrodynamics1932, mouginGeneralizedKirchhoffEquations2002} for a rigid body moving through an infinite, ideal fluid. They are valuable for evaluating fluid-structure coupling algorithms, like FloatStepper, because they are a set of ODE's that can be solved easily and fast on a computer and encompass non-trivial body-fluid interaction. In fact, for 3--dimensional motion they exhibit chaotic motion \cite{arefChaoticMotionSolid1993}.

If we restrict ourselves to motion in the infinite $xy$-plane and choose coordinate axes such that $S = 0$ (always possible because of symmetry of added mass matrix), then \eqn{eq:KKforce} simplifies to
\begin{subequations}\label{eq:2DForce}
\begin{align}
	\vek F_h & = T(\bs \omega\times\vek v_0) - \bs\omega \times T\vek v_0 - T\dot{\vek v}_0 \\
	\bs \tau_{h,0} & = -J\dot{\bs\omega} -\vek v_0\times T\vek v_0, \label{eq:tauh02D}
\end{align}
\end{subequations}
where now $\vek F_h = (F_x,F_y,0)^T$, $\vek v_0 = (v_x,v_y,0)^T$, $\bs \omega = (0,0,\omega)^T$ and $\bs \tau_{h,0} = (0,0,\tau)^T$. For such planar motion, the equations are integrable, but still exhibit interesting dynamics and coupling between the three degrees of freedom. From \eqn{eq:2DForce} we see that the hydrodynamic force depends both on body velocity and on body acceleration, when $\bs\omega \not= \vek 0$. Also it shows (last term on right hand side of \ref{eq:tauh02D}) that the torque with respect to $\vek x_0$ depends on velocity, even if there is no body rotation.

Many useful coupling validation cases can be constructed based on the Kirchhoff equations and their solutions. Here we consider a body of elliptic shape with major and minor axes $R(1+a^2)$ and $R(1-a^2)$, respectively, where $a\in[0,1]$. For such a body the added mass coefficients are,
\begin{equation}\label{eq:A_ii}
	A_{11} = \rho_f \pi R^2 (1-a^2)^2, \quad A_{22} = \rho_f \pi R^2 (1+a^2)^2, \quad 
	A_{66} = 2 \rho_f \pi R^4a^4.
\end{equation}
In our numerical experiment we use $R = 1$ m, $a = 0.5$, fluid density $\rho_f = 1$ kg/m$^3$ and body density $\rho_b = 0$ kg/m$^3$. The body is initialised with its centre at the origin and its minor axis aligned with the $x$-axis. The initial velocity is chosen to be $v_x = 1$ m/s, $v_y = 0$ m/s and $\omega = 1$ rad/s, which gives rise to an undulatory motion along the $x$-axis while the body wiggles with an angular amplitude of around 11$^\circ$. Our simulation is started at $t = 0$ s and ends at time $t = 6$ s corresponding to around $4.5$ motion periods. Adaptive time stepping was used with a maximum Courant-–Friedrichs–-Lewy (CFL) number of 0.1. The circular outer rim is placed at $40R$. Simulations were done with three different mesh resolutions with $100$, $200$ and $400$ cells in the radial direction with grading such that the inner most cells are 50 times smaller than the outer most cells. The corresponding azimuthal resolutions were $120$, $240$ and $480$ cells. Simulations obtained with the three mesh resolutions are shown in \fig{fig:wigglingTrajs} together with the exact solution obtained by integrating the Kirchhoff equations directly. \fig{fig:wigglingCoord} shows the horizontal, vertical and angular body components and their convergence to the exact solution (black curves) with mesh refinement. The added mass coefficients are recalculated in each time step. \fig{fig:wigglingMadd} shows the relative error in the added mass coefficients as a function of time with respect to \eqn{eq:A_ii}. The added mass is very close to constant throughout each simulation, and the error is seen to diminish with increased mesh resolution. As in the rising circle case it is possible that some of this error is due to the finite domain size.

\subsection{Freely floating box in regular waves}
\begin{figure}[tb]
    \centering
		\includegraphics[width=\textwidth]{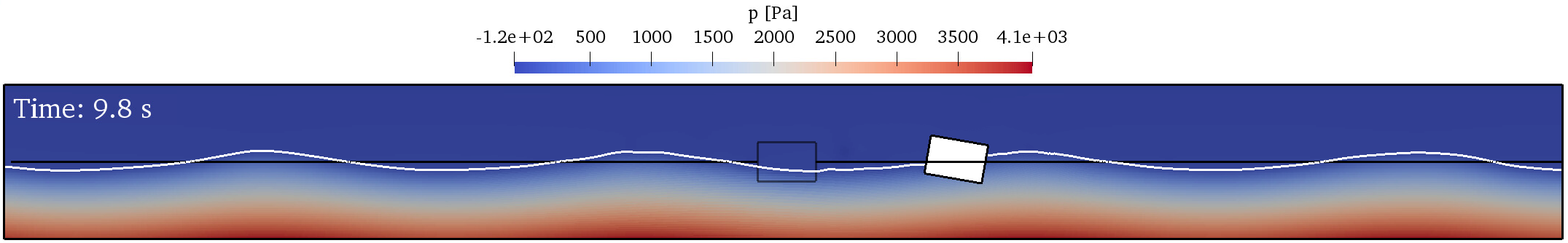}
    	\caption{Snapshots of water surface and box position at time $t = 0$ s (black) and $t = 9.8$ s (white) for the $H = 0.10$ m case of Ren et al. \cite{renNonlinearSimulationsWaveinduced2015}.\label{fig:Ren2015H004CaseIllustration}}
\end{figure}

We now increase the level of complexity by considering a case combining a free surface with several active degrees of freedom. We choose the benchmark case presented in Ren et al. \cite{renNonlinearSimulationsWaveinduced2015} with a box floating freely in regular waves in a wave flume. Here we try to reproduce their experimental data, including recorded wave elevation and box surge ($x$), heave ($y$) and pitch ($\theta$) motion. The physical wave flume is 23 m long, 44 cm wide and filled to a water depth of $d = 40$ cm. The floating box is 30 cm long, 20 cm high and 40 cm wide, leaving a gap of 2 cm to each side wall of the flume. The box is made of 8 mm thick Perspex plates and has a compartment at the middle which is filled with a granular material to give it an overall density of 500 kg/m$^3$, while retaining its centre of mass at its geometric centre. The total mass of the floater is then 12 kg.  Assuming the density of Perspex to be 1180 kg/m$^3$, and that the granular filling material is evenly distributed in the inner cross sectional area of the box, we calculate the moment of inertia with respect to its long centre axis to be $I_\textrm{box} = 0.151$ kg m$^2$. The box is initialised in equilibrium, half immersed in water ($\rho_w = 1000$ kg/m$^3$), with its centre 2 m from a piston-type wave generating wall placed at one end of the flume. To minimize wave reflections, a wave absorber is placed in the other end of the flume. Ren et al. \cite{renNonlinearSimulationsWaveinduced2015} perform two free floater tests with regular waves of wave height $H = 0.04$ m and $H = 0.10$ m, respectively, both with wave period $T = 1.2$ s. No detailed information is provided about the type of waves produced. In our numerical setup we generate waves using a custom made piston--type wave maker. It works by squeezing and stretching the cells in a region in front of the wave piston wall such as to make the piston wall move in accordance with a user supplied displacement file. The piston displacement signal we use is given from wave piston theory by 
\begin{equation}
	X(t) = -H\frac{\sinh(kd)\cosh(kd)+kd}{4\sinh^2(kd)}\sin(\omega t),
\end{equation}
where $\omega = 2\pi/T$, and $k$ is found numerically by solving the dispersion relation,
\begin{equation}
	\omega^2 =  g k \tanh(kd),
\end{equation}
for the given choice of period, depth and gravity, $g$. Given the two--dimensional nature of the problem with only a small gap between the box and flume walls, we choose to model the experiment in a two--dimensional setup. We choose a domain height of 0.8 m with the initial horizontal water surface in the middle at $y = 0$ m. To limit the mesh size, we truncate the flume 8 m from the wave--piston wall and use the active wave absorption boundary condition build into OpenFOAM to minimise wave reflections \cite{higueraRealisticWaveGeneration2013, higueraSimulatingCoastalEngineering2013}. The cells in our coarse base mesh are squares with side length 1 cm,  giving a mesh size of 63.400 cells. We also use a fine mesh with the region between $y = 0.2$ m and $y = 0.55$ m (covering the water surface and the box) refined into squares of side length $0.5$ cm giving a total of 145.600 cells. The simulations are run on the 16 cores of an AMD EPYC 7301 processor with time steps adjusted to keep the maximum CFL number in the domain below 0.5. On the coarse mesh numerical experiments were also done with CFL $<$ 0.1. Snapshots of the $H = 0.10$ m case simulation are shown in \fig{fig:Ren2015H004CaseIllustration}.

\begin{figure}[tb]
    \centering
    \begin{subfigure}[b]{0.48\textwidth}
    	\centering
	    \includegraphics[width=\textwidth]{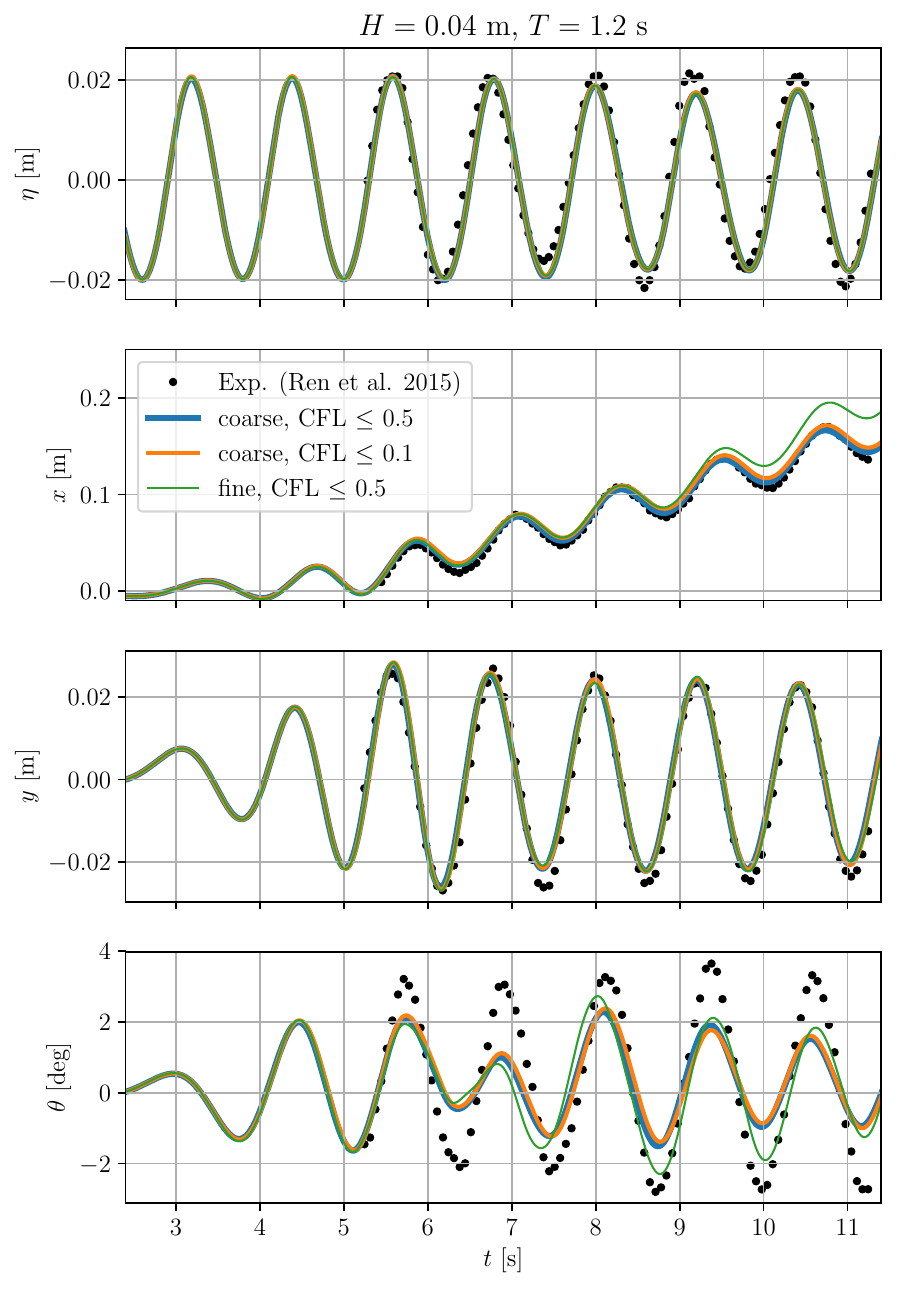}
	    \caption{\label{fig:Ren2015H004}}
	\end{subfigure}
    \begin{subfigure}[b]{0.48\textwidth}
    	\centering
	    \includegraphics[width=\textwidth]{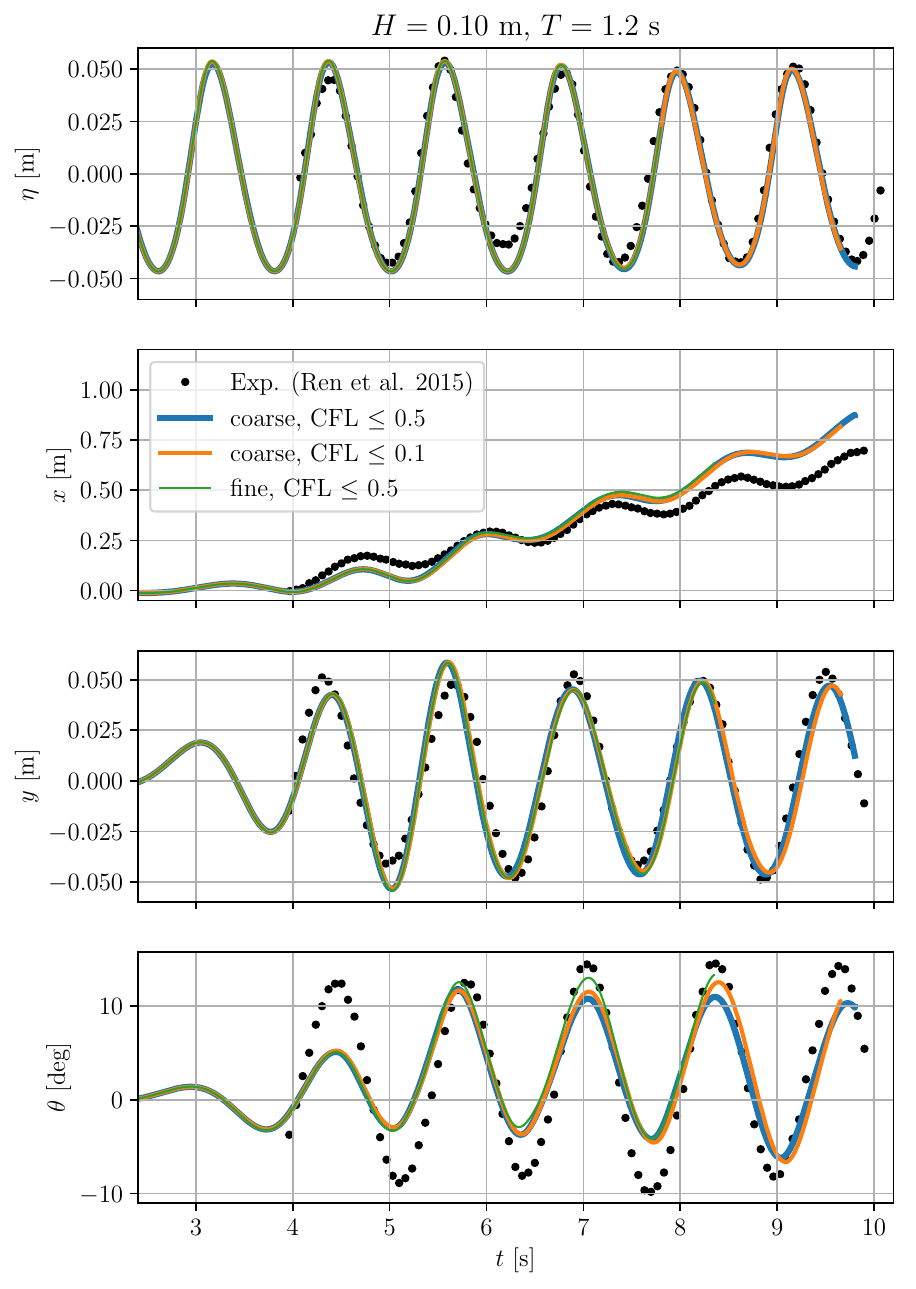}
	    \caption{\label{fig:Ren2015H010}}
	\end{subfigure}
    \caption{Freely floating box exposed to regular waves. Experimental results from Ren et al. \cite{renNonlinearSimulationsWaveinduced2015} reproduced with FloatStepper using different mesh and time resolution.\label{fig:Ren2015vsFloatStepper}}
\end{figure}

Ren et al. \cite{renNonlinearSimulationsWaveinduced2015} provide experimental data for five wave periods on the interval $t\in[0,\ 6]$ s. This data is plotted with black dots in \fig{fig:Ren2015vsFloatStepper}, and shows almost periodic motion with only minor differences in amplitude from one period to another. No data or information is given about the preceding time interval where the waves and body motion was building up. Unless the box was kept fixed during this ramp up period, it will have drifted some distance from its initial position at $x = 2$ m. There is therefore some uncertainty about the offset for the surge motion shown in the second row of \fig{fig:Ren2015vsFloatStepper}. In our numerical experiments we have observed that this distance may be of importance because the waves reflected from the box interacts with the incoming waves in the region between wave piston and box. Thus, if we start our box at $x=2$ m we generally overestimate the surge drift. Dom\'inguez et al. \cite{dominguez2019sph} have also tried to numerically reproduce the results in \cite{renNonlinearSimulationsWaveinduced2015}, and have successfully reproduced the surge drift in their highest resolution simulation. They do not explicitly mention their starting position for the box, but from their figures, we infer that they started it at $x = 4$ m. We will therefore use this starting position in our simulations. Our FloatStepper results for the two mesh and time resolutions are shown in \fig{fig:Ren2015vsFloatStepper}.

For the free surface elevation in the top row, Ren et al. \cite{renNonlinearSimulationsWaveinduced2015} do not mention where in the domain it is recorded. We have chosen to record the surface elevation just in front of the wave piston at $x = 0.3$ m. We compensate for the the phase difference caused by the different numerical and experimental wave gauge positions by shifting the experimental wave data by $4.4T$ along the time axis for the $H = 0.04$m case, and by $3.4T$ for the $H = 0.10$ m case. The results demonstrate a good match in wave height and period between the experimental data and all three numerical runs. 

The second row of panels in \fig{fig:Ren2015vsFloatStepper} shows the box surge motion which is characterised by oscillations superimposed on a steady drift in the direction of wave propagation. For the $H = 0.04$ case the coarse simulations with CFL = 0.1 and 0.5 capture both oscillations and drift very well. The fine simulation overestimates the drift for the last two periods. We do not currently have a good explanation for this behaviour. For the $H = 0.10$ m case all three simulations give virtually identical surge, but with an overestimation of the drift motion in all five wave periods. As mentioned, this may be due to differences in horizontal initial box position. We note that the accumulated drift of around 0.7 m during the five wave periods causes significant mesh distortion with the currently available SLERP based mesh deformation in OpenFOAM. The two coarse meshed simulations crash at around $t = 10$ s due to this, and the fine meshed simulation crashes at $t = 8$ s. In future work we will incorporate an improved mesh deformation method that allows for larger lateral displacement without compromising mesh quality. We also plan to couple FloatStepper with the overset mesh implementation in OpenFOAM, which will allow for arbitrarily large body displacements and rotations without the problem of deteriorating mesh quality.

The heave motion of the box is shown in the third row of \fig{fig:Ren2015vsFloatStepper}. This is captured very well for both wave heights with virtually no difference between the simulations with different mesh and time resolution.

We show the pitch motion of the body in the last row of \fig{fig:Ren2015vsFloatStepper}. For both wave heights, the experimental pitch data is characterized by oscillations with minor irregularities in amplitude. All our simulations underestimate the amplitude of the pitch oscillations. For the $H = 0.04$ case, the numerical oscillations are quite irregular and with minor, but noticeable difference between the coarse and fine simulations. For the $H = 0.10$ m case the simulated pitch oscillations are more regular, but with a slightly longer period than the experiments. This may be due to the overestimated numerical surge drift, causing a Doppler--like shift in the period of the pitch forcing from the incoming waves.

Taking the uncertainties about wave signal, ramp--up procedure, wave gauge position and initial body position into account, we regard the numerical results as satisfactory.

\subsection{Moored floating box in regular waves}

\begin{figure}[tb]
    \centering
   \includegraphics[width=0.5\textwidth]{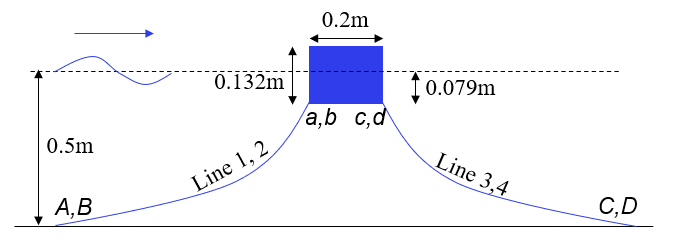}
    \caption{Numerical setup for the moored floating box experiment \cite{wuCFDSimulationPassively2018}, including the fairlead and anchor points denoted as [a, b, c, d] and [A, B, C, D], respectively. The coordinates for these points are provided in the Table \ref{tab:MooredBoxParameters}. \label{fig:Wu2019vsFloatStepperSetup}}
\end{figure}

\begin{table}[]
\centering
\begin{tabular}{lc}
\hline
\multicolumn{1}{c}{\textbf{Parameter}} & \textbf{Value}                    \\ \hline
Box length                             & 0.2 m                              \\
Box width                              & 0.2 m                              \\
Box height                             & 0.132 m                            \\
Box weight                             & 3.148 kg      \\
Centre of gravity                      & (0,0,-0.0126) \\
Box draft                              & 0.0786 m     \\ \hline
Mooring diameter                       & 0.003656                                 \\
Mooring weight                         & 0.0607 kg/m                              \\
Mooring length                         & 1.455 m                                  \\
Axial Stiffness                        & 29 N                                 \\
Fairlead a,b,c,d                       & $\pm$0.1,\ $\pm$0.1,\ -0.0736  \\
Anchor A,B,C,D                         & $\pm$1.385,\ $\pm$0.423,\ -0.5 \\ \hline
\end{tabular}
\caption{Box and mooring parameters along with coordinates of the mooring line anchor and fairlead connections from the experiment \cite{wuCFDSimulationPassively2018} }
\label{tab:MooredBoxParameters}
\end{table}

\begin{figure}[tb]
    \centering
   \includegraphics[width=0.45\textwidth]{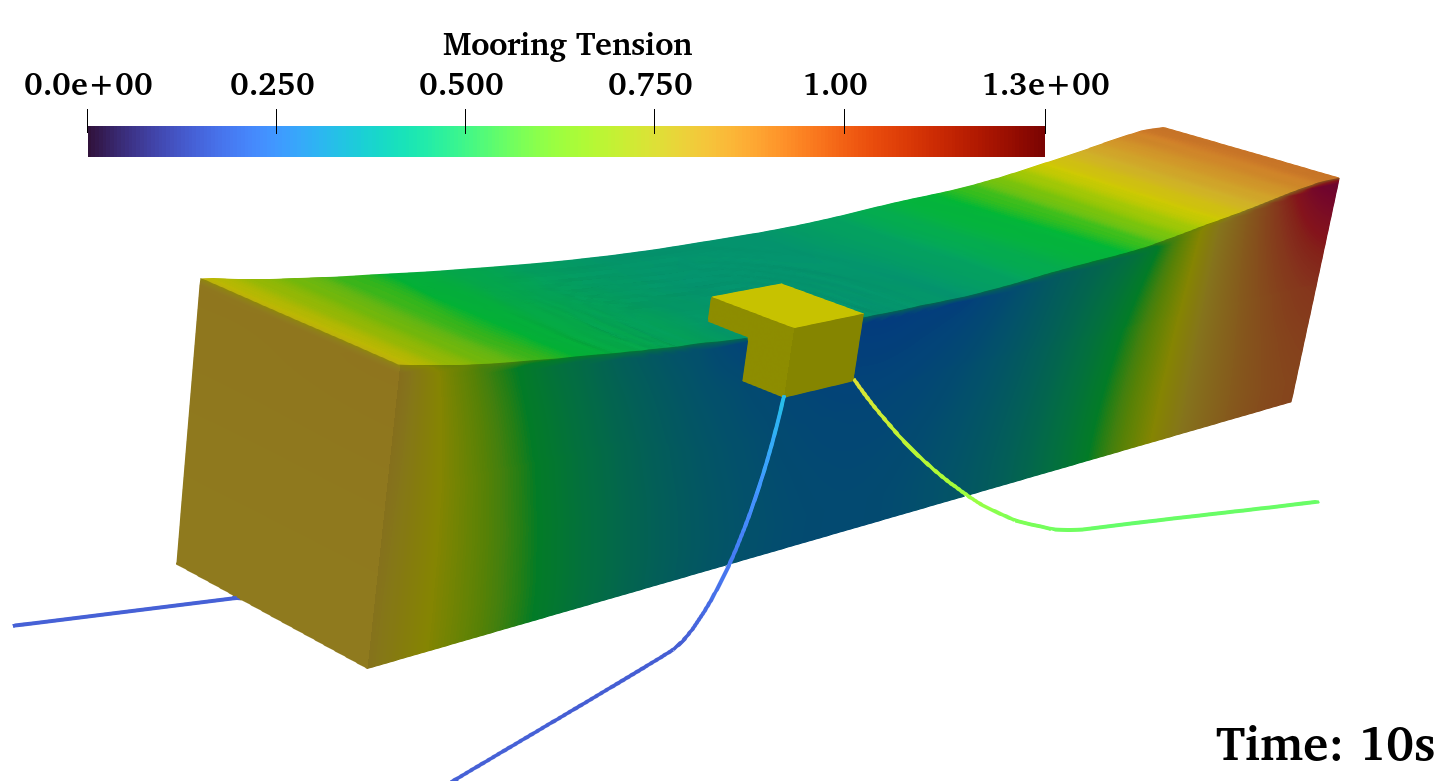}
   \includegraphics[width=0.45\textwidth]{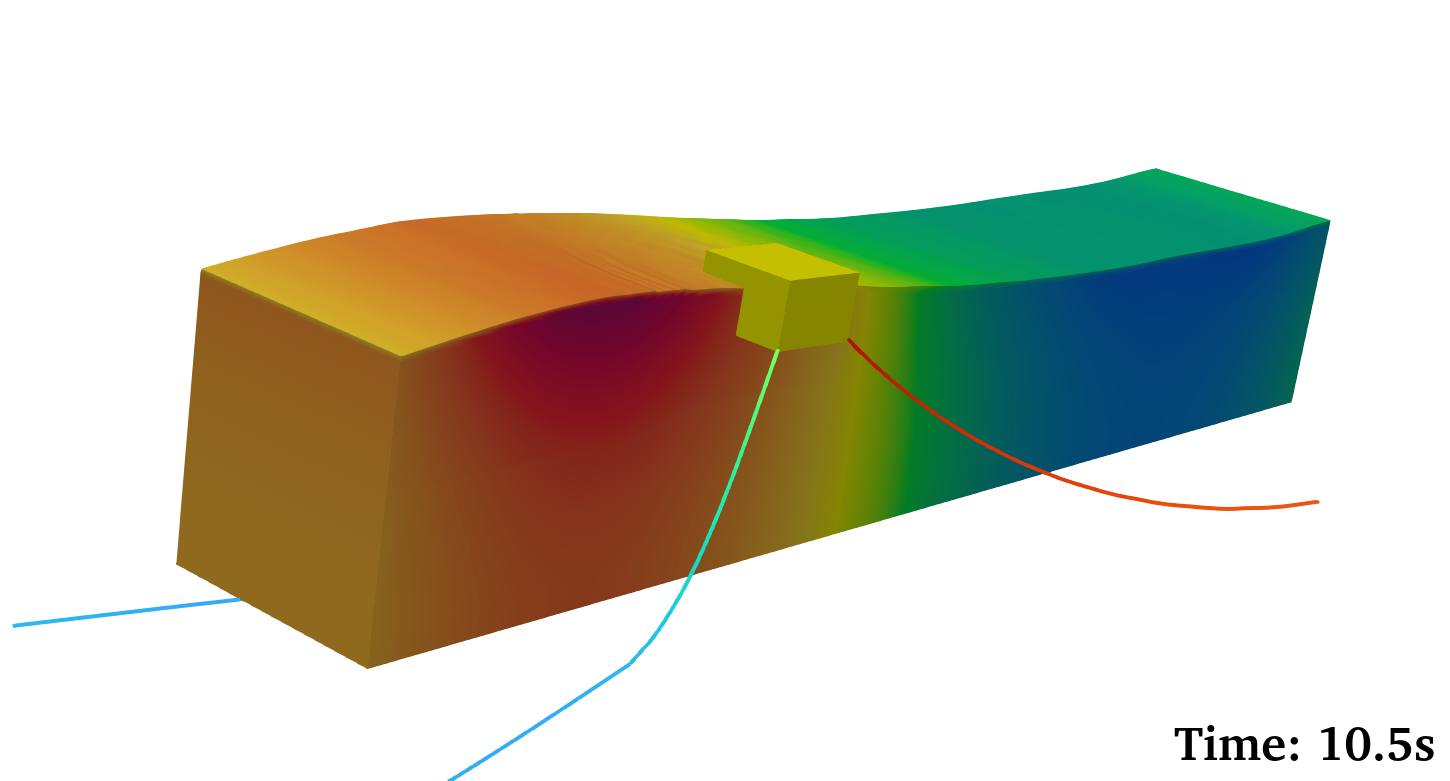}
   \includegraphics[width=0.45\textwidth]{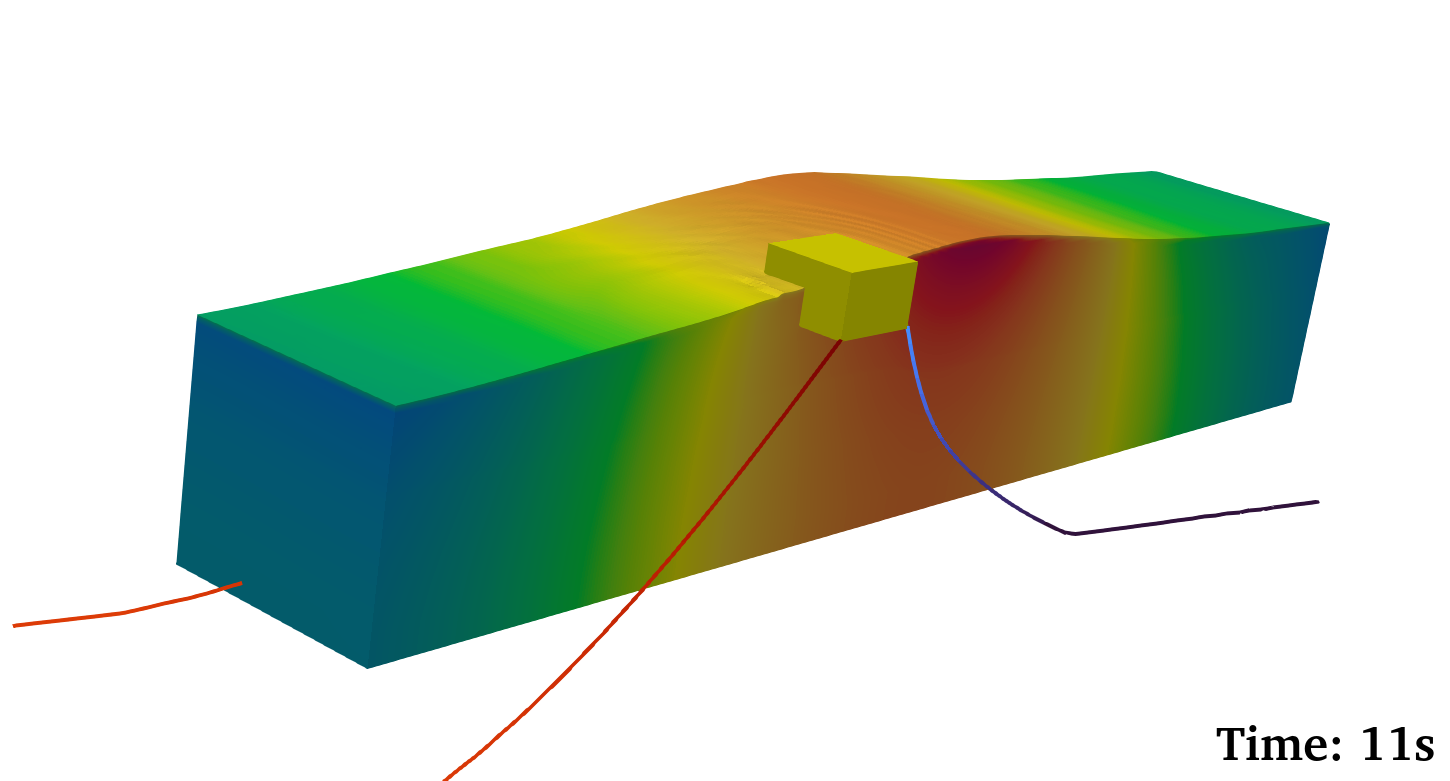}
   \includegraphics[width=0.45\textwidth]{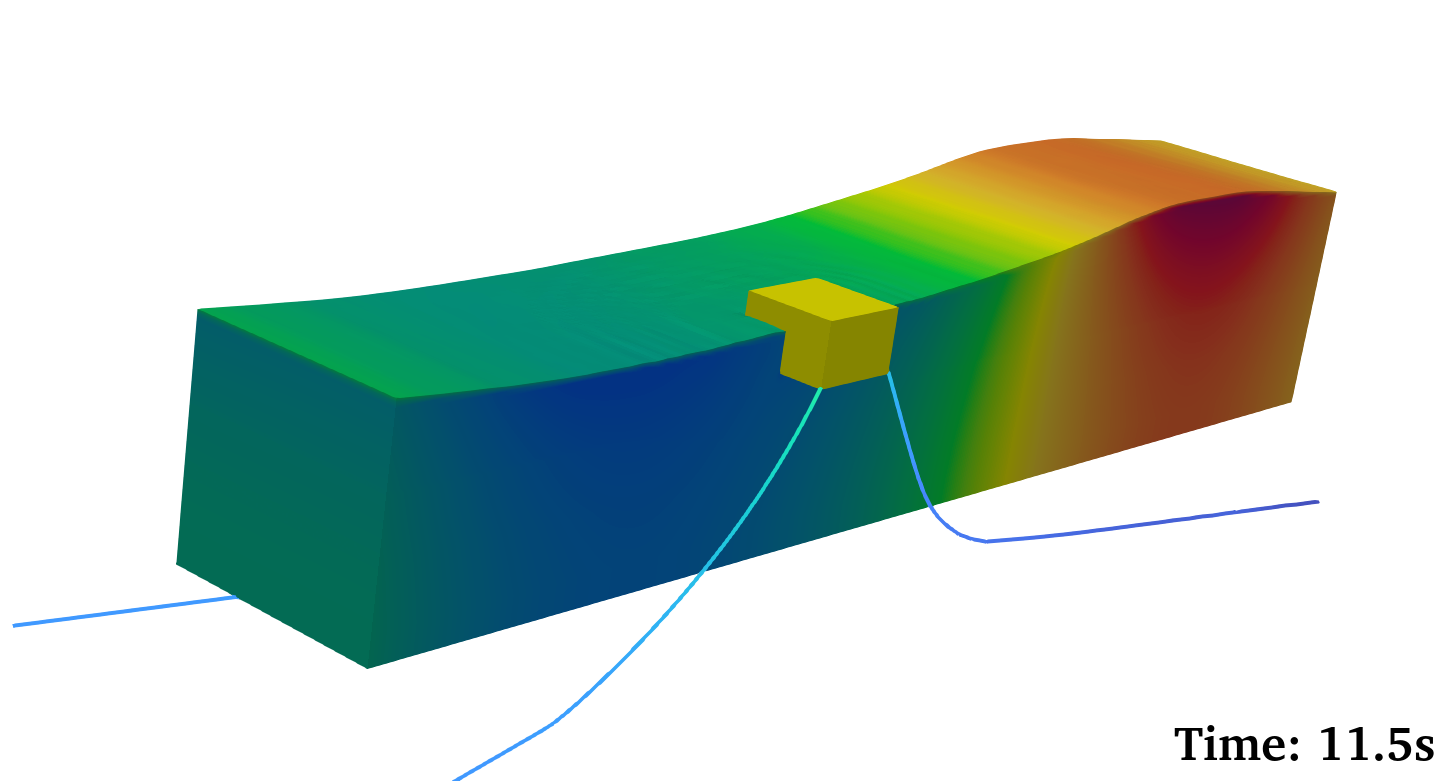}
    \caption{Instantaneous pressure distribution and a catenary mooring system configuration over one wave cycle for the box interacting with regular waves ($T = 2$ s, $H = 0.12$ m).
    \label{fig:MooredBoxAnimation}}
\end{figure}

\begin{figure}[tb]
    \centering
   \includegraphics[width=\textwidth]{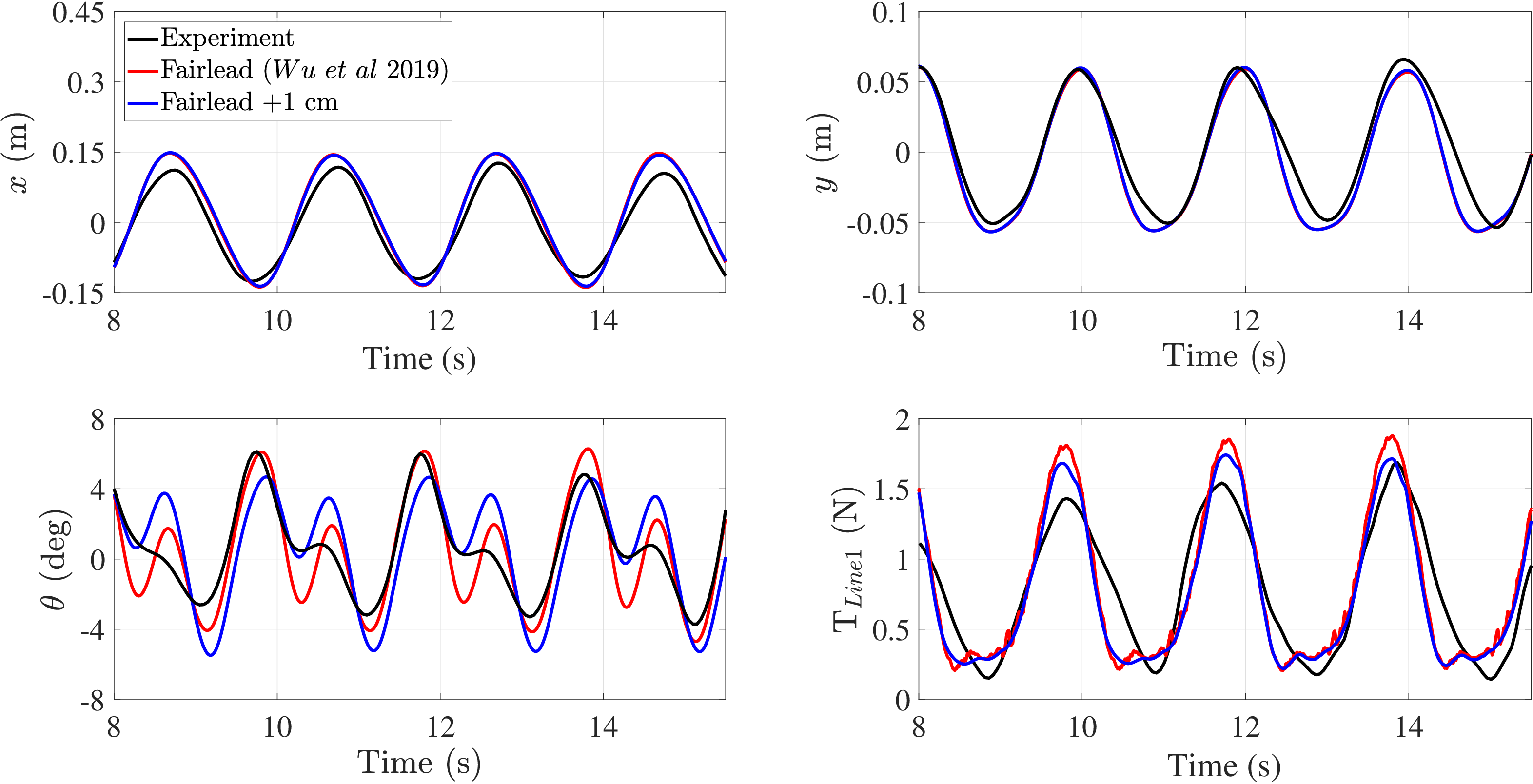}
    \caption{Moored floating box exposed to regular wave with $T = 2$ s and $H = 0.12$ m. CFD results with fairlead attachment as described in \cite{wuCFDSimulationPassively2018} (red) and with fairlead coordinate displaced by 1 cm (blue).\label{fig:Wu2019vsFloatStepper}}
\end{figure}

Our last test case focuses on a moored floating box in regular waves. In order to enhance the usability of FloatStepper in Marine and Offshore Engineering, we have employed an integrated approach that combines the solver with state of the art mooring line algorithms. Many approaches to mooring dynamics have been explored throughout the years such as MDD \cite{dewey1999mooring}, OrcaFlex \cite{randolph2009non}, Moody \cite{ferri2015implementation}, and MoorDyn \cite{hall2015validation}. It has been demonstrated that for floating structures subjected to significant motions, the behaviors described by a dynamic mooring model as opposed to a quasi-static mooring model produce a notable difference in mooring tensions \cite{aliyar2022numerical}. In the present work, FloatStepper is coupled with the dynamic mooring line tool MoorDyn, which is an open-source library designed to seamlessly integrate with rigid body solvers. This comprehensive library encompasses catenary moorings, seabed friction, axial and bending stiffness, hydrodynamic drag and added mass effects, making it a versatile and complete mooring model. The coupling between the FloatStepper and MoorDyn follows a similar methodology as that adopted in \cite{aliyar2022numerical}. The FloatStepper conveys the floating body's position and velocity to MoorDyn. MoorDyn, in turn, computes fairlead kinematics and updates the mooring system states, including the position, velocity of mooring line nodes, and segment tension. Subsequently, MoorDyn returns the mooring restraining forces and moments, which result from the sum of all fairlead tensions. These values are sent back from MoorDyn to the body motion solver, facilitating the update of the body's acceleration. MoorDyn's fundamental routines are built into a shared library to allow for the dynamic loading of functions within the FloatStepper.

The coupled model is validated against experimental measurements \cite{wuCFDSimulationPassively2018} involving a floating box moored using four catenary lines and exposed to regular wave conditions. The physical tests were conducted in a 30 m long and 1 m wide wave flume with a water depth of 0.5 m. Box dimensions and mooring configuration is shown in \fig{fig:Wu2019vsFloatStepperSetup} and listed in Table \ref{tab:MooredBoxParameters}. The box is constructed from lightweight PVC material, with a net density of 570 kg/m$^3$. A wooden plate was attached to the front of the box, positioned to face the incoming waves. Light reflecting markers were strategically positioned on this plate to track the 6-DoF motion. The mooring system comprises four chains of equal length arranged in a catenary configuration. To reduce the computational cost, we shorten our the numerical wave flume to 10 m. The domain is 1 m high and 1 m wide covered by a mesh of approximately 5 million cells. The mesh was decomposed into 100 sub-domains and simulations were executed on an AMD EPYC based HPC cluster.

Three different cases with differing wave conditions are found in the related references (\cite{dominguez2019sph}, \cite{chen2022cfd}, \cite{jeon2023moored}). We have run all three cases and found similar degree of correspondence between experiments and FloatStepper results in all of them. Here, we therefore only show the case with the most severe wave condition with a wave length of 4.116 m, a wave height of 0.12 m, and a wave period of 2 s (Case 3 from \cite{dominguez2019sph}, Case 2 from \cite{chen2022cfd}). The case was run for 20 simulation seconds corresponding to 10 wave periods with a maximum CFL number limit of 0.5. No turbulence model was activated. This approach simplified mesh generation and reduced computational effort, as it eliminated the need for strong grid refinement in the body boundary layers. Different turbulence models were tested in \cite{chen2022cfd} for the same test case and it was found that the choice of turbulence model had minimal impact on the results. \fig{fig:MooredBoxAnimation} depicts snapshots during a single wave period, illustrating the instantaneous conditions of the fluid domain and mooring lines. In this representation, the colour of the fluid domain corresponds to the dynamic pressure variation, and the colour of the mooring lines indicates the real-time tension within each segment of the line. 

\fig{fig:Wu2019vsFloatStepper} presents a comparison of the floating box's motion, focusing on surge ($x$), heave ($y$), pitch ($\theta$) and fairlead tension of Line 1 ($T_{Line 1}$). While our predictions for heave motion exhibit satisfactory agreement, there is a slight overestimation in surge motion and mooring tension $T_{Line1}$ amplitudes. Such discrepancies were also found in previous numerical results \cite{chen2022cfd} and \cite{dominguez2019sph} reported. In terms of pitch motion, our simulation matches maxima, minima, and phase, although there is a trend to overestimate the secondary peak amplitude between the main peaks, a phenomenon also observed in other numerical studies \cite{chen2022cfd, dominguez2019sph, jeon2023moored}. In these studies, the pitch discrepancies were attributed to uncertainties in experimental measurements, variations in geometry definitions, and the absence of the wooden plate for camera markers in the numerical simulations, which may alter the floater mass properties. In our current investigation, we analysed the effect of changing the center of gravity, inertia, and mooring parameters, to better understand the observed discrepancies. Notably, shifting the fairlead point just 1 cm closer to the center of gravity had a significant impact on the behaviour of the secondary peaks in pitch motion, as shown in \fig{fig:Wu2019vsFloatStepper}. We have also noticed a similar level of sensitivity when examining other wave case scenarios, mainly in the context of secondary peaks in the pitch motion. Given the reported uncertainties in experimental parameters, and the sensitivity of the rotational motion to these, we regard the moored floating body test as an acceptable validation result for FloatStepper.

\section{Summary and discussion}

We have demonstrated the feasibility of a new coupling algorithm, FloatStepper, for FVM based CFD simulation of an incompressible fluid and a rigid body. The method is based on direct calculation of the instantaneous added mass matrix, which allows separation of the added mass force from the other hydrodynamic forces. Hereby the equations of motion can be solved robustly without iteration. While other researchers have proposed to introduce explicit added mass calculation, the combination of  direct evaluation, non-iterative form and accessibility in a widely used open source CFD software framework is a novelty of our work.  

The robustness of the algorithm has been demonstrated through five simple test cases. First, for a rising disc in unbounded fluid, the solver is able to determine the acceleration with a relative error of less than 0.01\%. Next, for a disk falling into flat water, the solver is able to handle the abrupt change in added mass by a factor of $\rho_w/\rho_a = 830$ at the initial entry and we demonstrated convergence with mesh refinement. The hydrodynamic coupling between translational and rotational degrees of freedom was tested against a benchmark with a wiggling ellipse travelling through unbounded fluid, where the Kirchhoff equations provide an exact solution. For this test, a zero body mass was used to demonstrate the absence of added mass instability, and the solver was shown to converge to the analytical solution upon mesh refinement. 

The solver performance for floating structures in waves was next benchmarked in two test cases with a box floating freely and exposed to regular waves. Body motion was found to be reasonably well predicted with only small dependency on mesh and time resolution but with some overestimation of surge drift for the case with largest wave height and underestimation of pitch amplitude for both cases. Some of the observed discrepancy may be ascribed to the lack of exhaustive description of the experimental setup details such as wave and floater behaviour prior to the recorded experimental time interval.

Our last test case included a coupling with the MoorDyn library to compute the motion of a moored box in regular waves. The surge was well matched while the pitch motion deviated through a larger amplitude of a secondary motion peak in between the main peaks. The mooring line tension amplitude was overestimated by around 15\%. For this experiment, some uncertainty for the mass properties of the setup has been described by other researchers, who have discussed similar deviations. Although a closer match in pitch for both box cases is desirable, we note the good match for the ellipse case and also good matches for comparison to test results for a wind turbine floater presented at the Wind Energy Science Conference 2023 \cite{zenodoAnalysisInstallation}. On this basis, we regard the present test cases as an acceptable validation of the new algorithm. 

The current implementation in OpenFOAM is published as open source \cite{roenby_2023_8146516} in the hope that it will be used an extended by the CFD community, scientists and engineers working with floating objects. The shared code is at a proof-of-concept level of maturity. This means that, while it can certainly be used for production CFD runs for floating object simulations, there is still plenty of room for improvement. In particular, the code can most likely be optimized in terms of speed and memory usage. A central aspect here is the explicit added mass calculation. This comes at the price of solving six Poisson equations in the full domain, but eliminates the need for outer corrections, and -- just as importantly -- the uncertainty associated with choosing a safe, yet efficient, value for the acceleration relaxation parameter. In principle, FloatStepper with 6 active DoFs will cost the same as running with eight outer corrector (one zero--acceleration time step and six added mass column calculations in addition to the real time step). However, our experience so far is, that an added mass column calculation is not nearly as expensive as the full PISO time step of an outer corrector iteration. A quantification of this computational cost difference will be the subject of further studies, where we will also investigate the effect on accuracy and efficiency of reducing the added mass updating frequency.

Several numerical aspects, such as the ODE solver for the 6--DoF update, the interface advection method, and the type of mesh morphing, are currently hardcoded in the FloatStepper implementation. Our plan is to extend the code to allow the user various choices of schemes and methods and to easily add and test own customised methods. An important future extension would be to couple the method with overset mesh and immersed boundary methods to allow more extreme body motions than what is feasible with the deforming mesh method. Another relevant extension area would be the ability to handle multiple rigid bodies, which would enable simulations e.g. of the interaction of an installation vessel with a floating offshore wind turbine foundation.

A robust floating body algorithm is a prerequisite for realising the full potential of CFD as an engineering tool within fluid-structure interaction. It is our hope that the open source release of FloatStepper will help realise this potential, and foster collaboration in the CFD community to further improve the predictive capabilities of floating body CFD. 

\vskip1pc

\section*{Acknowledgement}
The work presented here was funded by the FloatStep Grand Solution project (8055-00075B) from Innovation Fund Denmark to Stromning Aps and Technical University of Denmark. JR also acknowledges partial funding from the DFF Sapere Aude Research Leader grant, InterFlow, to Roskilde University by Independent Research Fund Denmark (9063-00018B). JR thanks Henning Scheufler for useful discussions about code structure, and \v{Z}eljko Tukovi\'c for useful discussions about pressure boundary conditions.

\pagebreak

\vskip2pc



\bibliographystyle{RS}

\bibliography{FloatStepper}

\end{document}